\documentclass[11pt,a4paper]{article}

\usepackage[T1]{fontenc}
\usepackage{lmodern}
\usepackage[margin=25mm]{geometry}
\usepackage{amsmath,amssymb,booktabs,bm,placeins}
\usepackage{graphicx,array,tabularx,multirow}
\usepackage{microtype,cite}
\usepackage[font=small,labelfont=bf]{caption}
\usepackage[hidelinks]{hyperref}
\hypersetup{
 pdftitle={A reachability criterion for spectrally factorable SPDC in coupled waveguides without periodic poling},
 pdfauthor={Qiang Sun, Ke Wang, Andrew D. Greentree}
}

\newcommand{\rmd}{\mathrm{d}}
\newcommand{\ii}{\mathrm{i}}

\newcommand{\DevicePurity}{0.789}
\newcommand{\DeviceExactPurity}{0.797}
\newcommand{\DeviceFirstOrderPurity}{0.806}
\newcommand{\DeviceHalfWindow}{12}
\newcommand{\DeviceEdgeProbability}{1.31\times10^{-4}}
\newcommand{\DeviceSchmidt}{1.267}
\newcommand{\DevicePumpFWHM}{0.375}
\newcommand{\DeviceGamma}{2.77\times10^{-3}}
\newcommand{\DeviceMismatch}{4.710}
\newcommand{\DeviceDs}{427.126}
\newcommand{\DeviceDi}{-384.145}
\newcommand{\DeviceRho}{-0.899}
\newcommand{\DeviceHalfSplit}{2.25\times10^{4}}
\newcommand{\DeviceIdlerShare}{0.632}

\newcommand{\SIDevicePurity}{0.789}
\newcommand{\SIDeviceExactPurity}{0.797}
\newcommand{\SIDeviceFirstOrderPurity}{0.806}
\newcommand{\SIDeviceHalfWindow}{12}
\newcommand{\SIDeviceEdgeProbability}{1.31\times10^{-4}}
\newcommand{\SIDevicePumpFWHM}{0.375}
\newcommand{\SIDeviceGamma}{2.77\times10^{-3}}
\newcommand{\SIDeviceNeffP}{1.909}
\newcommand{\SIDeviceNeffS}{1.960}
\newcommand{\SIDeviceNeffI}{1.833}
\newcommand{\SIDeviceMismatch}{4.710}
\newcommand{\SIDeviceNgP}{2.327}
\newcommand{\SIDeviceNgS}{2.199}
\newcommand{\SIDeviceNgI}{2.442}
\newcommand{\SIDeviceDs}{427.126}
\newcommand{\SIDeviceDi}{-384.145}
\newcommand{\SIDeviceRho}{-0.899}
\newcommand{\SIDeviceSlope}{42.981}
\newcommand{\SIDeviceHalfSplit}{2.25\times10^{4}}
\newcommand{\SIDeviceIdlerShare}{0.632}
\newcommand{\SIDeviceSchmidt}{1.267}

\begin{document}

\title{A reachability criterion for spectrally factorable SPDC in coupled
waveguides without periodic poling}

\author{Qiang Sun\textsuperscript{1,2,*}, Ke Wang\textsuperscript{1}, and
Andrew D. Greentree\textsuperscript{2}}
\date{}
\maketitle
\begin{center}
\small
\textsuperscript{1}Department of Electrical and Electronic Engineering,\\
School of Engineering, RMIT University, Melbourne, Victoria 3001, Australia\\[4pt]
\textsuperscript{2}Department of Physics, School of Science,\\
RMIT University, Melbourne, Victoria 3001, Australia\\[4pt]
\textsuperscript{*}\href{mailto:qiang.sun@rmit.edu.au}{qiang.sun@rmit.edu.au}
\end{center}

\begin{abstract}
Producing spectrally pure photon pairs by spontaneous parametric
down-conversion without periodic poling would broaden the choice of nonlinear
materials.  It requires optical modes that provide both phase matching and
suitable group velocities.  We derive a joint reachability criterion for a
constant-cross-section coupler with a nonlinear main guide and a linear
auxiliary.  In a two-mode description, phase closure fixes the hybridization
parameter that also determines the main-guide participation and the attainable
group-delay ratio.  This relation screens material and mode pairs and specifies
detuning and participation targets for full-vector refinement.  A first-order
purity map connects the reachable delay ratio and bandwidth--length product to
the photon-pair spectrum.  A full-vector AlN/TiO$_2$ feasibility example
satisfies direct phase matching with a ratio of pump--idler to pump--signal
group-delay mismatches of $-0.899$ and yields a calculated unfiltered Schmidt purity of 0.789
over 5~mm.  The criterion links phase closure, dispersion control, and
nonlinear participation when assessing coupled-waveguide sources beyond
periodically poled materials.
\end{abstract}

\section{Introduction}

Spontaneous parametric down-conversion (SPDC) supplies correlated photon pairs
and heralded single photons for quantum photonics
~\cite{Burnham1970,Couteau2018,Zhang2021}.  Quantum networks and photonic
quantum computing often require photons from independent sources to interfere
~\cite{Knill2001,Kimble2008}, making their spectral--temporal mode quality
important alongside pair generation.  Frequency correlations in the joint spectral
amplitude (JSA) leave a heralded photon in a mixed state; the Schmidt purity $P$
quantifies its single-mode character and, for identical independent sources
with ideal heralding and no other distinguishability, sets the
Hong--Ou--Mandel visibility
~\cite{Grice1997,Grice1998,Law2000,URen2005,Mosley2008}.
Filtering can raise the purity at the cost of pair transmission and heralding
efficiency~\cite{Christ2012,MeyerScott2017}.  Designing the generated spectrum
to be nearly factorable avoids that trade-off.

Spectral factorability requires control of both phase matching and the
group-velocity relation.  Central phase matching makes generation
amplitudes add coherently along the device, while group velocities determine
the orientation of the phase-matching function relative to the pump spectrum.
Birefringence and waveguide dispersion can supply both conditions
~\cite{URen2005,Mosley2008}; related dispersion engineering is also used in
four-wave-mixing sources~\cite{GarayPalmett2007}.  Quasi-phase matching (QPM)
adds a separate momentum degree of freedom~\cite{Armstrong1962,Fejer1992}.
In periodically poled QPM, repeated domain inversion supplies a reciprocal
vector, allowing a geometry chosen for its group velocities to retain central
phase matching through the poling period.  This separation of tasks underlies
the widespread use of periodic poling in lithium-niobate nonlinear photonics
~\cite{Wang2018,Zhu2021,Boes2023}.  QPM also extends beyond ferroelectric
domain inversion: all-optical poling in dispersion-engineered Si$_3$N$_4$
waveguides combines an induced nonlinear grating with group-velocity matching
for broadband second-harmonic generation~\cite{Nitiss2020BroadbandQPM}.

Ferroelectric periodic poling constrains the generating material and its fabrication.
It requires controlled domain inversion over the interaction length, and
longitudinal nonuniformity alters the phase-matching spectrum
~\cite{Zhao2023}.  Segmented micro-heaters can compensate thickness-induced
spectral distortion in thin-film PPLN~\cite{Li2024}.  The domain-inversion
requirement does not transfer directly to conventional
AlN or SiC platforms~\cite{Suceava2023,Weiss2025NoPoling}.  Liquids introduce a
different setting: a recent ferroelectric-nematic SPDC demonstration used a
molecular twist to provide reconfigurable QPM
~\cite{Sultanov2024LiquidSPDC}.  A route without a longitudinal grating would
allow nonlinear materials to be selected without requiring periodic inversion
of their quadratic response.  The difficulty is then to satisfy phase and
group-velocity matching within the same modal structure.

Coupled-waveguide sources already offer ways to meet this difficulty.
Stationary transverse supermodes can satisfy direct nonlinear phase matching
~\cite{Liu2019}, distinct from coupling-length QPM based on longitudinal power
beating~\cite{Biaggio2014,Barral2019CouplingQPM}.  Main \textit{et al.}
combined a nonlinear lithium-niobate guide with a linear silica microfiber:
hybridization of the pump mode provided direct phase matching, group-velocity
control, and a full-vector JSA calculation~\cite{Main2016Hybrid}.  Their
asymmetric-coupler analysis subsequently used a hybridized daughter mode and
its coupling dispersion to reshape the photon-pair spectrum
~\cite{Main2019PRA}.  Heterogeneous couplers have also been designed for
group-velocity matching with QPM~\cite{Ding2021Heterogeneous}.  Separating
nonlinear and linear-material roles has precedent in classical conversion
~\cite{Luo2019SemiNonlinear}, and the role of frequency-dependent coupling is
established~\cite{Shi2023}.  The present analysis builds on these ingredients
to identify the constraint linking phase closure, delay control, and nonlinear
participation.

The remaining challenge is to determine whether phase matching and a target
group-delay ratio can be achieved on the same stationary hybrid branch before
full-vector optimization.  These conditions are coupled: the hybridization
that closes the phase mismatch also fixes the modal participation and the
group slowness.  A useful design criterion must therefore evaluate both
requirements at the same operating point.

The analysis in this work has three linked contributions.
\textit{(i)} Our main result is a two-mode joint phase--delay reachability
criterion: phase closure fixes the participation at which the required delay
correction must occur.  It distinguishes a delay-only screen from the joint
condition on one branch and guides the subsequent full-vector refinement.
\textit{(ii)} As the spectral consequence, we map reachable delay ratios to a
scale-free first-order purity landscape, connecting the modal target to the
photon-pair spectrum.
\textit{(iii)} A full-vector AlN/TiO$_2$ feasibility design shows that carrier
closure, opposite-sign delays, and a nonzero nonlinear overlap can coexist in
one cross-section.  Supporting derivations, numerical checks, and alternative
solid and liquid examples are collected in Appendices~\ref{sec:S_purity}--\ref{sec:S_phasecheck}.

\section{Joint phase--delay reachability}
\label{sec:theory}

\subsection{A stationary supermode supplies both corrections}

\begin{figure}[t]
 \centering
 \includegraphics[width=0.95\textwidth]
 {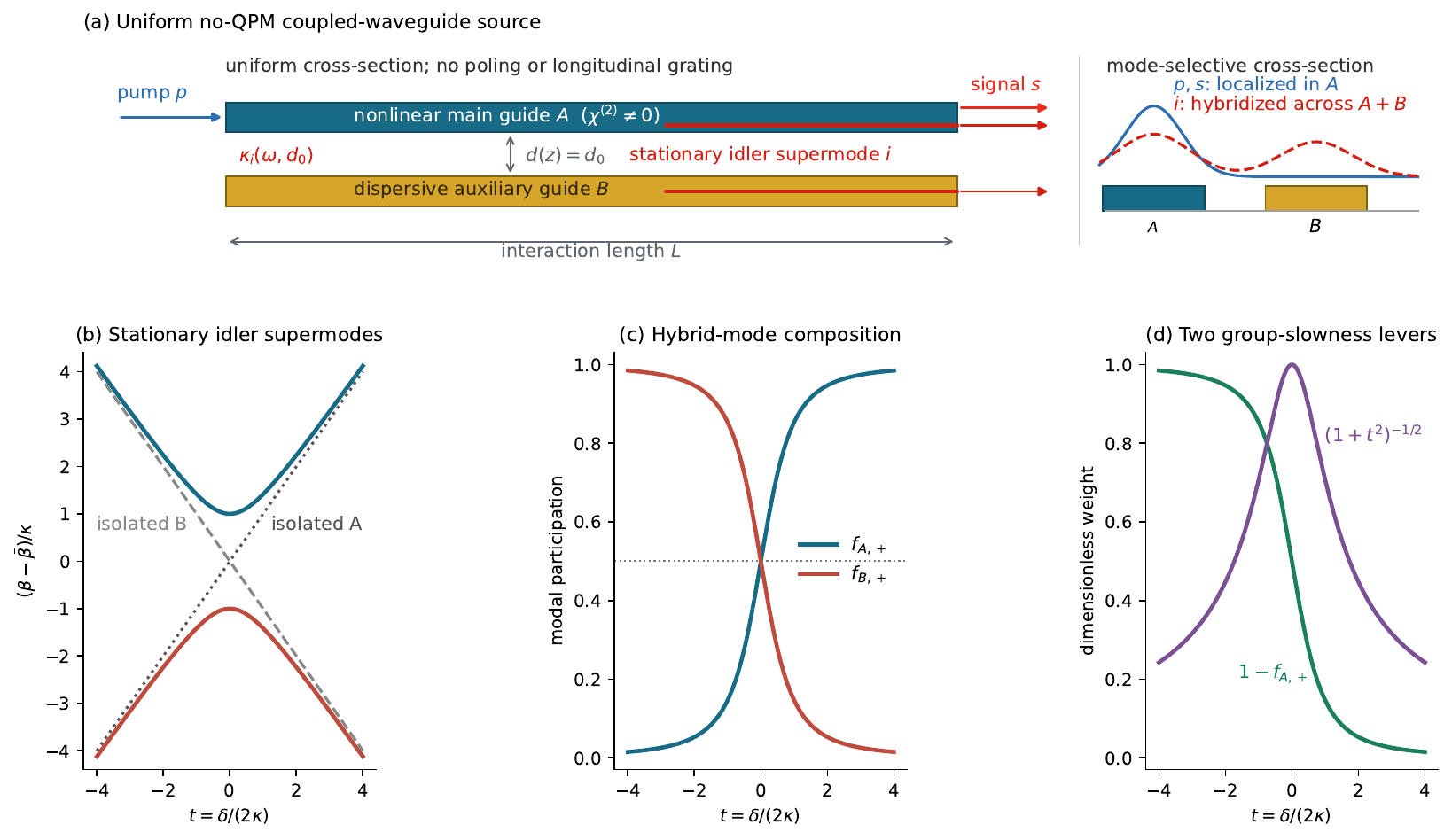}
 \caption{Stationary hybridization links phase and delay control.
 (a) Nonlinear main guide $A$ and dispersive auxiliary $B$ at constant
 separation; pump and signal are predominantly in $A$, while the idler is a
 supermode.  (b) An avoided crossing shifts the idler propagation constant.
 (c) The same detuning fixes its main-guide participation.
 (d) Participation-weighted group slowness and explicit coupling dispersion
 determine its delay.  These are linked controls: once phase matching selects
 a point on a branch in (b), the participation and delay weights in (c,d)
 must be evaluated at that same point.}
 \label{fig:mechanism}
\end{figure}

Figure~\ref{fig:mechanism}(a) defines the system: a nonlinear main guide $A$
and a linear auxiliary $B$ retain a constant cross-section and separation
$d(z)=d_0$ over a length $L$.  The pump and signal remain predominantly in $A$;
the idler occupies a stationary supermode spanning both guides.  The model
does not use longitudinal power beating as a grating.  Relabeling the fields
gives the corresponding construction for another hybridized mode.

Let $\omega_\mu$ and $\beta_\mu$ denote carrier angular frequencies and modal
propagation constants, with $\mu=p,s,i$ for pump, signal, and idler.  Energy
conservation requires $\omega_p=\omega_s+\omega_i$ but does not select absolute
wavelengths.  A prime denotes differentiation with respect to angular
frequency at fixed geometry, evaluated at the corresponding carrier:
$\beta'_\mu\equiv[\partial\beta_\mu(\omega)/\partial\omega]_{\omega=\omega_\mu}$.
This derivative is the modal group slowness, $\beta'_\mu=1/v_{g,\mu}$,
where $v_{g,\mu}$ is the group velocity.  The relevant conditions are
\begin{equation}
\begin{aligned}
 \Delta k_0&=\beta_p-\beta_s-\beta_i=0,\\
 D_s&=\beta'_p-\beta'_s,\qquad D_i=\beta'_p-\beta'_i,\qquad
 \rho=\frac{D_i}{D_s}.
\end{aligned}
\label{eq:targets}
\end{equation}
Here $D_s$ and $D_i$ are pump--daughter group-delay mismatches per unit length.
We assume $D_s\ne0$; otherwise the daughter labels can be interchanged or the
unnormalized delays used.  Without a QPM vector, the modal propagation
constants must supply both phase closure and a useful value of $\rho$.

In an orthonormal two-mode basis, take the coupling to be real and positive.
The standard idler matrix and its eigenvalues are
~\cite{Liu2019,Main2019PRA,Ding2021Heterogeneous,Shi2023}
\begin{equation}
 \mathbf H_i(\omega,d)=
 \begin{pmatrix}
  \beta_{iA}(\omega)&\kappa_i(\omega,d)\\
  \kappa_i(\omega,d)&\beta_{iB}(\omega)
 \end{pmatrix},
 \label{eq:H}
\end{equation}
\begin{equation}
 \beta_{i,\sigma}=\frac{\beta_{iA}+\beta_{iB}}{2}
 +\sigma\sqrt{\left(\frac{\beta_{iA}-\beta_{iB}}{2}\right)^2+\kappa_i^2},
 \qquad \sigma=\pm1.
 \label{eq:supermode}
\end{equation}
The splitting in Fig.~\ref{fig:mechanism}(b) is $2\kappa_i$ at isolated-mode
resonance.  Moving along either branch changes both its propagation constant
and transverse composition.  Direct phase matching selects the value
\begin{equation}
 \beta_{i,\sigma}=\beta_*\equiv\beta_{pA}-\beta_{sA}.
 \label{eq:directpm}
\end{equation}
The pump and signal are taken to be unchanged from their main-guide modes in
this reduction.  In a complete device their propagation constants are
calculated from the full cross-section.

If $f_A$ is the main-guide basis weight of the selected idler eigenvector,
differentiating Eq.~\eqref{eq:supermode} at the idler carrier gives
\begin{equation}
 \beta'_{i,\sigma}=f_A\beta'_{iA}+(1-f_A)\beta'_{iB}
 +\sigma\frac{\kappa_i\kappa_i'}
 {\sqrt{[(\beta_{iA}-\beta_{iB})/2]^2+\kappa_i^2}}.
\label{eq:groupvelocity}
\end{equation}
Here $\kappa_i'\equiv[\partial\kappa_i(\omega,d)/\partial\omega]_{\omega=\omega_i}$
is the coupling dispersion at fixed separation $d$ and fixed cross-section.
The derivatives $\beta'_{iA}$, $\beta'_{iB}$, and $\beta'_{i,\sigma}$
are the group slownesses of the two isolated idler modes and the selected
idler supermode, respectively, at that same carrier.
Figures~\ref{fig:mechanism}(c,d) show why participation alone is insufficient
to determine the delay.  The material-weighted term changes with $f_A$, while
the explicit coupling-dispersion term is concentrated near the avoided
crossing.  Phase matching selects a point on this same branch; it cannot
choose either contribution independently.

\subsection{Stage 1: the delay-only screen}

Define the isolated-main-guide delays and normalized contrasts by
\begin{equation}
\begin{aligned}
 \tau_s&=\beta'_{pA}-\beta'_{sA},&
 \tau_i&=\beta'_{pA}-\beta'_{iA},\\
 r&=\frac{\tau_i}{\tau_s},&
 m&=\frac{\beta'_{iB}-\beta'_{iA}}{\tau_s},\\
 q&=\frac{\kappa_i'}{\tau_s},&
 t&=\frac{\beta_{iA}-\beta_{iB}}{2\kappa_i}.
\end{aligned}
\label{eq:dimensionless}
\end{equation}
Since $D_s=\tau_s$ in this model, Eq.~\eqref{eq:groupvelocity} yields
\begin{equation}
 \rho=r-(1-f_A)m-\sigma\frac{q}{\sqrt{1+t^2}}.
 \label{eq:reachability}
\end{equation}
Thus $r$ is determined by the isolated generating guide, rather than assigned
as a free phenomenological delay ratio.  The auxiliary changes it through
participation, group-slowness contrast, and coupling dispersion.

For negligible explicit coupling dispersion, $q=0$, a target
$\rho_{\mathrm{tar}}$ requires
\begin{equation}
 f_A^{\mathrm{req}}=1-\frac{r-\rho_{\mathrm{tar}}}{m},
 \qquad 0\le f_A^{\mathrm{req}}\le1,
 \label{eq:requiredparticipation}
\end{equation}
for $m\ne0$.  If $m=q=0$, the auxiliary supplies no delay correction and
$\rho=r$.  Equation~\eqref{eq:requiredparticipation} tests whether sufficient
delay contrast exists, treating participation as adjustable.  Carrier phase
matching is imposed in Stage 2.

\begin{figure}[t]
 \centering
 \includegraphics[width=0.82\textwidth]
 {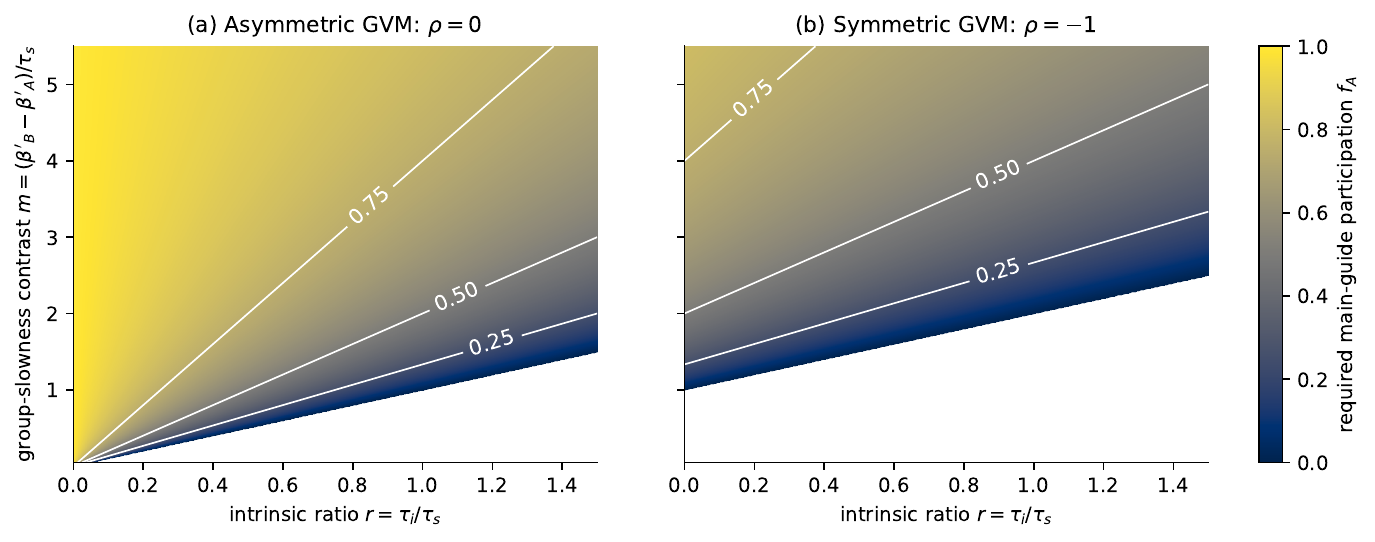}
 \caption{Stage 1: delay-only reachability for $q=0$.
 Color gives the required main-guide idler participation; contours mark
 $f_A=0.25$, 0.50, and 0.75 for (a) $\rho=0$ and (b) $\rho=-1$.
 White regions cannot supply the required delay correction at the displayed
 positive $r,m$ without a coupling-dispersion contribution.  A colored point
 is a candidate, not a phase-matched design: Stage 2 requires the
 participation selected by phase closure to equal that required here.}
 \label{fig:reachability}
\end{figure}

In Fig.~\ref{fig:reachability}(a), the asymmetric target $\rho=0$ becomes
accessible at $m=r$; the 50:50 contour is $m=2r$.  For the symmetric target
$\rho=-1$, Fig.~\ref{fig:reachability}(b) shifts those boundaries to $m=r+1$
and $m=2(1+r)$.  At a fixed positive $(r,m)$, the symmetric target therefore
requires a main-guide participation lower by $1/m$.  It demands a greater
transfer of modal weight to the auxiliary.  This comparison identifies a
delay--participation trade-off.  The second screen determines whether either participation
is compatible with the carrier condition.

\subsection{Stage 2: phase closure selects the allowed delay correction}
\label{sec:phaseclosure}

At the carriers, define the offsets of the required idler propagation
constant from the isolated modes:
\begin{equation}
 \Delta_A=\beta_*-\beta_{iA},\qquad
 \Delta_B=\beta_*-\beta_{iB},\qquad
 u=\frac{\Delta_A}{\kappa_i}.
 \label{eq:phaseoffsets}
\end{equation}
Evaluating the characteristic equation and normalized eigenvector of
Eq.~\eqref{eq:H} at $\beta_*$ gives the joint relation
\begin{equation}
 \boxed{\begin{aligned}
  \text{Phase closure:}\quad &\Delta_A\Delta_B=\kappa_i^2,\\
  \text{Main-guide participation:}\quad &f_A=\frac{1}{1+u^2},\\
  \text{Delay ratio:}\quad &\rho=r-\frac{m u^2+2qu}{1+u^2}.
 \end{aligned}}
 \label{eq:phaseclosure}
\end{equation}
\emph{Phase closure fixes $u$; the same $u$ fixes the participation and the
delay ratio.}  This is the constraint absent from the delay-only map.
The first line is necessary and sufficient for $\beta_*$ to be an eigenvalue
of the assumed two-mode matrix.  The second follows from the eigenvector
ratio $c_B/c_A=u$, and the third from its expectation value of
$\mathbf H_i'\equiv[\partial\mathbf H_i(\omega,d)/\partial\omega]_{\omega=\omega_i}$,
the elementwise frequency derivative at fixed geometry.  Appendix~\ref{sec:S_reachability} gives the derivation.

The product condition requires the offsets to have the same nonzero sign.
The target lies above both isolated modes on the upper branch ($u>0$), or
below both on the lower branch ($u<0$); a target between them is unreachable
with nonzero real coupling at those inputs.  In particular, a main guide
already phase matched in isolation, $\Delta_A=0$, cannot retain that same
phase-matched triplet after nonzero coupling without changing other inputs.
The generating guide generally needs to be detuned as well as the auxiliary.

For a prescribed delay ratio, let $a=r-\rho_{\mathrm{tar}}$.  The last line
of Eq.~\eqref{eq:phaseclosure} becomes
\begin{equation}
 (m-a)u^2+2qu-a=0.
 \label{eq:targetquadratic}
\end{equation}
A joint solution needs a real, nonzero root that also satisfies
$u=\Delta_A/\kappa_i$ and $\Delta_B=\kappa_i/u$.  Solving the quadratic
alone is therefore not sufficient.  For $q=0$, Eq.~\eqref{eq:requiredparticipation}
gives the delay-required weight; Eq.~\eqref{eq:phaseclosure} tests whether the
phase-matched branch has that weight.  With coupling dispersion retained,
the sign of $u$ also matters through $2qu$.

These relations identify candidate operating points within the two-mode
model.  Given isolated modes and a coupling model, they test carrier ordering
and delay targets and specify the detuning and participation to seek.
Geometry changes alter several inputs together.  Full-vector calculations
then refine these candidates and determine the accuracy of the reduction for
a physical cross-section.

\subsection{Participation also sets a generation cost}

If only $A$ is nonlinear, the coupling amplitude into the stationary modes is
\begin{equation}
 \Gamma=\mathcal C\int_A\!\rmd A
 \sum_{abc}\chi^{(2)}_{abc}e_{p,a}e^*_{s,b}e^*_{i,c},
 \label{eq:screeningoverlap}
\end{equation}
where the fields are power normalized and $\mathcal C$ collects the
normalization convention.  If hybridization mainly changes the idler
amplitude, the estimate
\begin{equation}
 |\Gamma|\simeq\sqrt{f_A}|\Gamma_A|,\qquad R/R_A\simeq f_A
 \label{eq:participationrate}
\end{equation}
relates the overlap and pair-rate scale to their main-guide values, with other
factors held fixed.  The participation required by phase and delay matching
therefore also has a generation cost.  This estimate is useful for screening;
the device example evaluates the vector integral directly.  If the auxiliary
is also nonlinear, its amplitude must be added coherently rather than
represented by Eq.~\eqref{eq:participationrate}.

\section{Spectral consequence: a scale-free first-order purity map}
\label{sec:purity-theory}

Once a phase-matched branch supplies $D_s$ and $D_i$, the first-order
spectral mismatch is
\begin{equation}
 \Delta k=D_s\nu_s+D_i\nu_i+O(\nu^2),
 \label{eq:slopes}
\end{equation}
where $\nu_s$ and $\nu_i$ are angular-frequency detunings.  For a Gaussian
pump with intensity-spectrum standard deviation $\sigma_p$, define
\begin{equation}
 x=\frac{\nu_s}{\sigma_p},\qquad y=\frac{\nu_i}{\sigma_p},\qquad
 B=\frac{L|D_s|\sigma_p}{2}.
 \label{eq:normalizedcoordinates}
\end{equation}
With a uniform longitudinal interaction and constant overlap, the JSA is,
up to separable spectral phases,
\begin{equation}
 F_\rho(x,y)\propto e^{-(x+y)^2/4}
 \operatorname{sinc}[B(x+\rho y)].
 \label{eq:normalizedjsa}
\end{equation}
Its normalized Schmidt weights $\lambda_n$ and purity $P=\sum_n\lambda_n^2$
depend only on $\rho$ and $B$~\cite{Law2000,Eberly2006}.
Thus the reachability calculation supplies $\rho$, and the pump bandwidth and
interaction length set $B$.  The map is scale free under these first-order,
Gaussian-pump, uniform-interaction assumptions.

\begin{figure}[t]
 \centering
 \includegraphics[width=0.94\textwidth]
 {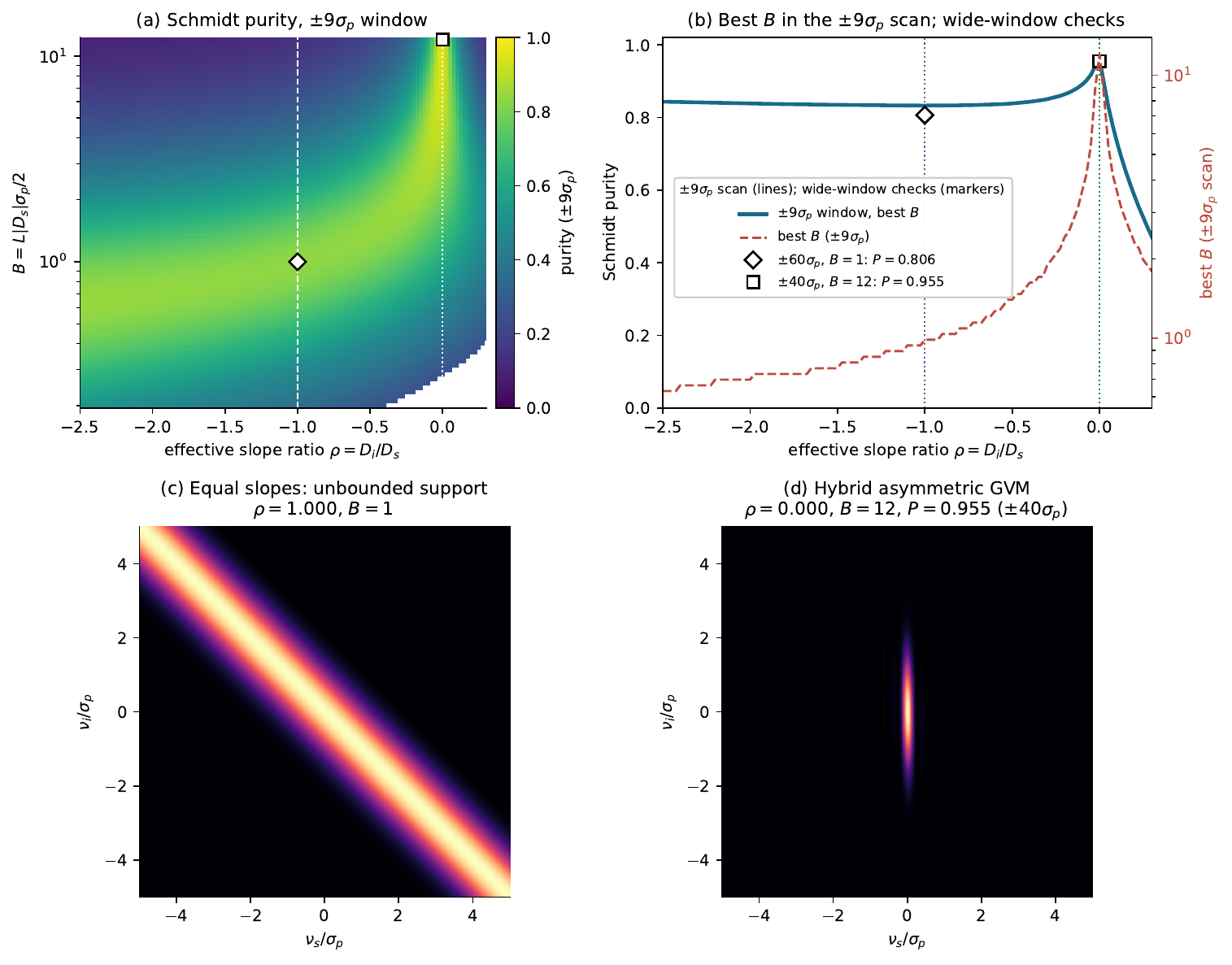}
 \caption{First-order spectral consequences of a reachable delay ratio.
 (a) Schmidt purity versus $\rho$ and $B$ over a $\pm9\sigma_p$ window;
 points with appreciable boundary probability are masked.
 (b) Purity at the best $B$ (solid) and the best $B$ (dashed, right axis)
 from the same scan; at $\rho=0$ the optimum lies at its edge, $B=12$.
 Separate wide-window markers give $P=0.806$ at $\rho=-1$, $B=1$ over
 $\pm60\sigma_p$ (diamond) and $P=0.955$ at $\rho=0$, $B=12$ over
 $\pm40\sigma_p$ (square).  They are not points on the finite-window curves,
 which give 0.833 and 0.960 at those parameters.
 (c) Equal slopes leave one spectral direction unconfined.
 (d) Asymmetric matching narrows one daughter coordinate without
 longitudinal apodization.}
 \label{fig:puritymap}
\end{figure}

Figure~\ref{fig:puritymap}(a) shows a high-purity ridge that moves with
$\rho$.  At symmetric matching, $\rho=-1$, the first-order phase-matching
axis is orthogonal to the pump axis.  At $B=1$, a wide-window calculation
gives $P=0.806$ over $\pm60\sigma_p$, decreasing to 0.804 over
$\pm120\sigma_p$ as weak sinc side lobes are included.  This familiar
near-0.81 benchmark belongs to a uniform interaction near symmetric matching,
not to coupled waveguides in general
~\cite{Graffitti2018Optica,Graffitti2018}.  The plotted $\pm9\sigma_p$ map
gives 0.833 at the same point because it excludes more of those tails.
Appendix~\ref{sec:S_purity} separates the numerical windows explicitly.

The ridge does not require exact symmetry.  In the plotted scan, the best
purity varies by less than 0.006 for $-2\le\rho\le-0.5$; wide-window
calculations optimized over $B$ give 0.814 at both endpoints.  Moving toward
$\rho=0$ shifts the favorable region to larger $B$, as shown in
Fig.~\ref{fig:puritymap}(b).  The tested asymmetric point, $\rho=0$, $B=12$,
gives $P=0.955$ over $\pm40\sigma_p$.  The different purities at symmetric
and asymmetric matching reflect different JSA orientations, rather than
competing estimates of one device.

Figures~\ref{fig:puritymap}(c,d) expose the distinction.  For $\rho=1$, both
factors depend only on $x+y$, leaving the orthogonal coordinate unconfined;
the first-order model does not then define a normalizable broadband state.
At $\rho=0$, phase matching confines only the signal coordinate, approaching
an axis-aligned spectrum for large $B$.  In a physical design, the vanished
idler linear coefficient makes higher-order dispersion particularly important.
A favorable first-order target must therefore still be checked using a JSA
calculated from the full modal dispersion.

Table~\ref{tab:benchmark_levels} distinguishes the theoretical benchmarks
from the material example that follows.  The Gaussian-kernel comparison is
an analytical extension, derived in Appendix~\ref{sec:S_gaussian} and discussed in
Sec.~\ref{sec:discussion}, separate from the constant-gap device calculation.

\begin{table}[t]
 \centering
 \caption{Purity values at distinct levels of the analysis.  Finite-window
 results retain the windows stated here and in Appendices~\ref{sec:S_purity} and~\ref{sec:S_jsa}.}
 \label{tab:benchmark_levels}
 \small
 \begin{tabular}{@{}p{0.26\textwidth}p{0.48\textwidth}l@{}}
  \toprule
  Level & Conditions & $P$\\
  \midrule
  Uniform, symmetric & First order; $\rho=-1$, $B=1$; $\pm60\sigma_p$ & 0.806\\
  Uniform, asymmetric & First order; $\rho=0$, $B=12$; $\pm40\sigma_p$ & 0.955\\
  Ideal Gaussian kernel & First order; matched Gaussian widths; untruncated & 1\\
  AlN/TiO$_2$ example & Modal dispersion and idler-dependent overlap;
  $\pm12$~THz & 0.789\\
  \bottomrule
 \end{tabular}
\end{table}

\section{AlN/\texorpdfstring{TiO$_2$}{TiO2}: a material-resolved feasibility example}
\label{sec:device}

\subsection{Numerical model}

The example asks whether a complete cross-section can close the carrier
mismatch, reach the near-symmetric delay regime, and retain a nonlinear
generation pathway.  AlN provides the quadratic response, while TiO$_2$ is
modeled as a linear dispersive auxiliary.  Deposition-based integrated
processes have been demonstrated separately for AlN and amorphous TiO$_2$
~\cite{Xiong2012AlN,Evans2015TiO2}, and unpoled modal nonlinear conversion in
AlN provides a relevant precedent~\cite{Guo2016}.  These results motivate
the material choice, without establishing a process for the combined
cross-section.

\begin{figure}[t]
 \centering
 \includegraphics[width=0.98\textwidth]
 {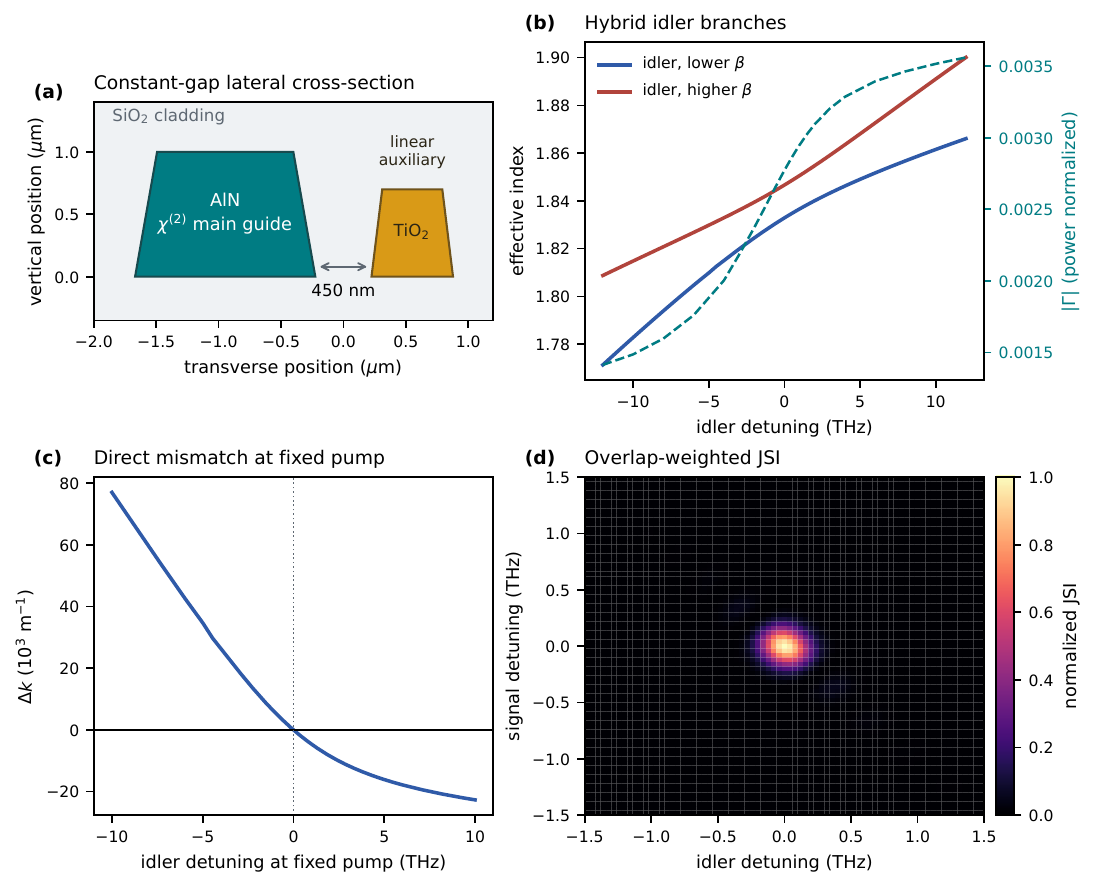}
 \caption{Full-vector feasibility example using a linear TiO$_2$ auxiliary.
 (a) Lateral cross-section; labels give top widths and bottom-edge gap.
 (b) Tracked idler effective indices and the magnitude of the AlN-only
 power-normalized overlap for the selected lower branch (dashed).
 (c) Direct mismatch along the fixed-pump energy-conservation line.
 (d) Overlap-weighted joint spectral intensity at the optimized pump bandwidth.
 The same stationary triplet supplies the carrier, delay, and spectral results.}
 \label{fig:devicevalidation}
\end{figure}

We solve a two-dimensional full-vector eigenproblem in COMSOL Multiphysics.
Figure~\ref{fig:devicevalidation}(a) shows silica-embedded lateral trapezoids
with a 450-nm bottom-edge gap.  The AlN guide is 1000~nm high and 1091~nm wide
at the top, with an $80^\circ$ sidewall angle; the TiO$_2$ guide is 700~nm
high and 481~nm wide, with an $83^\circ$ angle.  The AlN $c$ axis is normal to
the chip.  Geometry and material tensors are constant over the 5-mm
interaction: no periodic poling or longitudinal modulation is present.

The selected colors are 775, 1300, and 1919~nm.  We fixed the pump and signal
and obtained the idler from energy conservation; these wavelengths are inputs
to the material example, not to the scale-free theory.  An even higher-order
TM-like AlN pump and a fundamental TM-like AlN signal couple to the
lower-$\beta$ hybrid idler.  Selective excitation of the pump mode would be
required experimentally.  AlN and silica are dispersive; TiO$_2$ uses a
provisional isotropic Cauchy law.  Appendices~\ref{sec:S_model}--\ref{sec:S_jsa} document the
material functions, unrounded geometry, production mesh, and branch tracking.

\subsection{One triplet meets the phase and delay conditions}

The tracked idler branches in Fig.~\ref{fig:devicevalidation}(b) remain
separated while exchanging modal character through the hybridization region.
At the carrier, their half-splitting is
$\DeviceHalfSplit~\mathrm{m^{-1}}$.  The selected lower branch has an AlN
electric-field-intensity share of \DeviceIdlerShare{} within the two cores.
This core-domain diagnostic is distinct from the isolated-basis weight $f_A$
in Eq.~\eqref{eq:phaseclosure}.

Along the fixed-pump line in Fig.~\ref{fig:devicevalidation}(c), the mismatch
changes sign near the chosen carrier.  The retained numerical point gives
\begin{equation}
\begin{aligned}
 \Delta k_0&=\DeviceMismatch~\mathrm{m^{-1}},\\
 (D_s,D_i)&=(\DeviceDs,\DeviceDi)~\mathrm{ps/m},\\
 \rho&=\DeviceRho.
\end{aligned}
\label{eq:device_targets}
\end{equation}
Its 5-mm phase-matching factor
$\operatorname{sinc}^2(\Delta k_0L/2)$ rounds to 1.000.  The same
branch-tracked propagation constants give opposite-sign delays and a ratio
close to the symmetric target.  Both delays follow from the modal dispersion.

The dashed curve in Fig.~\ref{fig:devicevalidation}(b) tests the third part
of the design logic: sufficient field must remain in AlN to generate pairs.
We integrate the full vector tensor contraction over AlN only,
\begin{equation}
 \Gamma=\frac{\displaystyle\int_{\mathrm{AlN}}\!\rmd A
 \sum_{abc}d_{abc}E_{p,a}E^*_{s,b}E^*_{i,c}}
 {\sqrt{P_pP_sP_i}},\qquad
 P_\mu=\frac12\operatorname{Re}\int
 (\mathbf E_\mu\times\mathbf H_\mu^*)\cdot\hat{\mathbf z}\,\rmd A.
 \label{eq:gamma}
\end{equation}
Using $|d_{33}|=4.3$~pm/V, $|d_{31}|=|d_{33}|/45$
~\cite{Majkic2017}, and $|d_{15}|=|d_{31}|$ with a common nominal sign gives
$|\Gamma|=\DeviceGamma~\mathrm{V^2/W^{3/2}}$ at the carrier.
The $d_{33}$ term contributes 99.989\% of its magnitude.  TiO$_2$ is assigned
no quadratic response, so the AlN integral is the complete nonlinear channel
in this model.  Its finite value across the hybridization band confirms that
the generating pathway is retained as the idler delay changes.

\subsection{The spectral calculation follows from the selected branch}

For the constant-gap example we use
\begin{equation}
\begin{aligned}
 F(\omega_s,\omega_i)&=\mathcal N\,
 \Gamma(\omega_s,\omega_i)\alpha(\omega_s+\omega_i)\\
 &\quad\times\operatorname{sinc}\!\left[\frac{L\Delta k(\omega_s,\omega_i)}{2}\right]
 \exp\!\left[\frac{\ii L\Delta k(\omega_s,\omega_i)}{2}\right],
\end{aligned}
\label{eq:jsa}
\end{equation}
with $\Delta k$ from the branch-tracked propagation constants rather than a
first-order fit.  The numerical overlap weight uses $|\Gamma(\omega_i)|$:
the idler is recomputed across frequency, while pump and signal fields in the
overlap integral remain at their carrier solutions.  All three propagation
constants retain their frequency dependence.  Singular values $s_n$ of the
complex JSA give $\lambda_n=s_n^2/\sum_j s_j^2$, $P=\sum_n\lambda_n^2$,
and $K=1/P$~\cite{Law2000,Eberly2006}.

At $L=5$~mm, optimizing the Gaussian pump gives an intensity FWHM of
\DevicePumpFWHM~THz.  Three calculations at this same bandwidth and on the
same $\pm\DeviceHalfWindow$-THz window separate the contributions:
\begin{equation}
 P_{\mathrm{first\ order}}=\DeviceFirstOrderPurity,\qquad
 P_{\mathrm{modal},\,|\Gamma|=\mathrm{const.}}=\DeviceExactPurity,\qquad
 P_{\mathrm{modal},\,|\Gamma(\omega_i)|}=\DevicePurity.
 \label{eq:puritycontrols}
\end{equation}
No spectral filter is applied.  The outer grid cells contain probability
$\DeviceEdgeProbability$; the value is a calculated-window result, with the
window checks given in Appendix~\ref{sec:S_jsa}.  The reduction from 0.806 to 0.797
comes from higher-order modal dispersion, and the further reduction to 0.789
from the sampled overlap variation.  The departure of $\rho$ from $-1$ is
not the source of that reduction: the first-order control remains at 0.806.

\begin{figure}[t]
 \centering
 \includegraphics[width=0.88\textwidth]
 {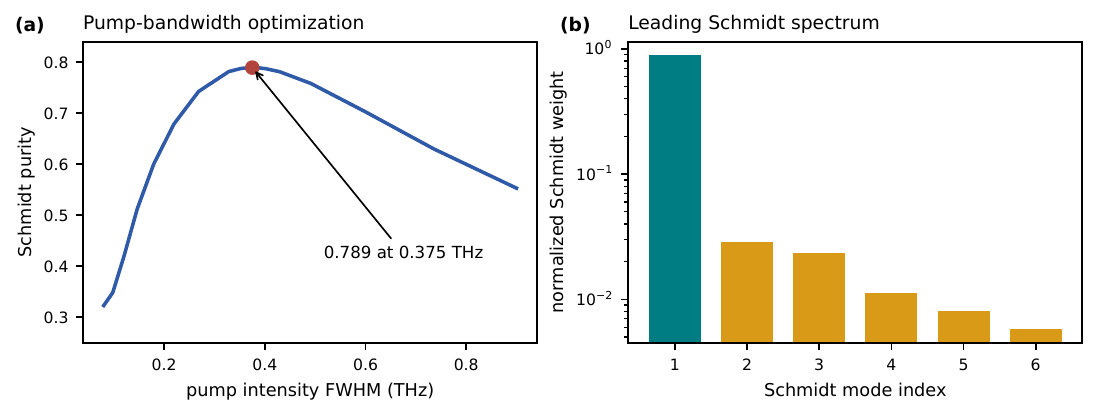}
 \caption{Schmidt analysis of the AlN-generated channel.
 (a) Purity versus pump intensity FWHM using modal dispersion and the sampled
 idler-dependent overlap magnitude.  The maximum is
 $P=\DevicePurity$ at \DevicePumpFWHM~THz.
 (b) Leading normalized Schmidt weights at that point.  Residual modes give
 $K=\DeviceSchmidt$; the joint intensity appears in Fig.~\ref{fig:devicevalidation}(d).}
 \label{fig:devicepurity}
\end{figure}

Figure~\ref{fig:devicevalidation}(d) shows the resulting compact central
joint spectrum and weaker sidelobes of the uniform interaction.  The panel
displays $|F|^2$; the Schmidt calculation also retains the propagation phase
in Eq.~\eqref{eq:jsa}.  Figure~\ref{fig:devicepurity}(a) resolves the
pump-bandwidth maximum: narrower pumps constrain the sum-frequency axis too
strongly, while broader pumps leave a larger phase-matching contribution to
the correlations.  At the optimum, the leading Schmidt weight in
Fig.~\ref{fig:devicepurity}(b) is 0.888; the residual weights give
$K=\DeviceSchmidt$.  These figures connect the reachable near-symmetric delay
ratio to its spectral consequence in the material example.

\subsection{Sensitivity and independent checks}
\label{sec:tolerance}

The joint target remains restrictive in a material-resolved cross-section.
Gap changes of $\pm30$~nm and top-width changes of $\pm20$~nm all reduce the
5-mm fixed-color phase-matching factor below 0.007.  Small width errors can
instead shift the generated colors: in the independent femwell calculation,
a $\pm1$-nm AlN width error moves the phase-matched signal by about
$\mp0.26$~THz and changes $\rho$ from $-0.903$ to $-0.975$ or $-0.830$.
Restoring phase matching therefore does not hold the delay ratio fixed,
consistent with the joint-constraint picture.

Material uncertainty gives a similar test.  Offsetting the provisional
TiO$_2$ index by $\delta n=\pm0.005$ and jointly retuning the two top widths
recovers $|\Delta k_0|<0.9~\mathrm{m^{-1}}$ and $\rho=-0.902$ at both
endpoints, with the gap unchanged.  These are design-stage corrections, not
post-fabrication tuning results.  Appendix~\ref{sec:S_sensitivity} gives the geometry,
index, and fixed-pump tests together.

An independent finite-element model agrees with the carrier effective indices
to within $6\times10^{-5}$ and gives $\rho=-0.903$ at its own phase-matched
point, compared with $-0.899$ in COMSOL (Appendix~\ref{sec:S_femwell}).  The separate
two-mode consistency check gives $u=-0.87$ on the lower branch and yields
$f_A=0.57$, against a projected weight of 0.57--0.58.  Its delay estimate is
$-0.88$, compared with $-0.91$ from the same solver's supermodes at nominal
colors.  Here coupling is inferred from the full-vector splitting, so this
agreement checks the reduction rather than independently predicting the
device.  All reported device phase, delay, and JSA values use the directly
calculated supermodes.

\section{Discussion}
\label{sec:discussion}

Equation~\eqref{eq:phaseclosure} places carrier phase matching, group-delay
correction, and nonlinear participation on the same hybrid branch.
Figure~\ref{fig:reachability} screens for sufficient delay contrast; phase
closure then selects the participation, and Fig.~\ref{fig:puritymap} gives
the first-order spectral consequence.  Within the two-mode model, this
sequence identifies compatible carrier ordering, required detuning,
participation, and delay targets, together with the associated nonlinear
generation trade-off.  These targets guide the subsequent full-vector search;
optimizing purity over independent delay parameters would miss their joint
constraint.

Direct supermode phase matching, hybridization-based
group-velocity control, and full-vector JSA calculations were already combined
in hybrid and coupled lithium-niobate sources
~\cite{Main2016Hybrid,Main2019PRA}.  Modal-phase-matched SPDC has also been
studied in other unpoled structures
~\cite{Weiss2025NoPoling,Shi2025SiCSPDC,Fang2025ModalSPDC,Ye2024BIC}.
The contribution here is the explicit condition tying these controls to the
same participation.  The AlN/TiO$_2$ example demonstrates the coexistence of
phase matching, opposite-sign group delays, and finite nonlinear overlap in
a full-vector waveguide model.

The division of roles expands the material combinations worth considering.
The generating guide supplies an effective quadratic response without
requiring periodic domain inversion.  The auxiliary can be selected for its
index and dispersion, independently of a quadratic response.  Appendices~\ref{sec:S_sic} and~\ref{sec:S_liquid} illustrate
this freedom with an alternative AlN/4H-SiC design and a preliminary
ferroelectric-nematic/Si$_3$N$_4$ example.  The latter uses a polar-ordered
liquid and demonstrates approximate carrier
closure and opposite-sign delays in the full modal calculation.  These
examples test transfer to other material combinations; the primary
quantitative example remains AlN/TiO$_2$.

The near-0.81 benchmark at symmetric group-velocity matching reflects the sinc
sidelobes of a uniform interaction, rather than a limit imposed by
hybridization.  An ideal Gaussian longitudinal generation kernel with matched
widths yields a separable spectrum in the first-order model
~\cite{QuesadaBranczyk2018,Branczyk2011,Dosseva2016,Tambasco2016,PovedaHospital2023}.
Gaussian-like effective nonlinearities have also been pursued through domain
engineering~\cite{Graffitti2018Optica,Pickston2021}.  Smooth, nonperiodic
variation of the waveguide separation offers a possible route to longitudinal
shaping.  It simultaneously changes the nonlinear overlap, phase mismatch,
and group delays, requiring a separate adiabatic inverse-design treatment.
Appendix~\ref{sec:S_gaussian} gives the Gaussian benchmark and the local-supermode
formulation for this prospective extension.

The criterion screens candidates within the assumed two-mode representation.
Its quantitative accuracy depends on the reduced matrix elements.  Independent
isolated-field coupled-mode calculations give a selected-branch
propagation-constant error of about $4.7\times10^2~\mathrm{m^{-1}}$, comparable
to the 5-mm half-maximum phase-mismatch acceptance of
$5.6\times10^2~\mathrm{m^{-1}}$ (Appendix~\ref{sec:S_phasecheck}).  Full-vector
refinement therefore remains essential.  The AlN/TiO$_2$ feasibility example
also relies on a provisional TiO$_2$ dispersion model and a finite-window
spectral calculation.  Practical implementation will require measured material
dispersion, fabrication-aware optimization, and device-level characterization
of loss, access-mode coupling, and photon-pair generation.

\section{Conclusion}

We have formulated a joint reachability criterion for coupled-waveguide SPDC
without periodic poling.  In the two-mode model, carrier phase closure fixes
the hybridization parameter, which in turn fixes the nonlinear-guide
participation and attainable group-delay ratio.  This links three quantities
that cannot be optimized independently.  The first-order purity map states
the spectral consequence of a reachable ratio, and a full-vector AlN/TiO$_2$
example meets the joint conditions with $\rho=\DeviceRho$ and a calculated
unfiltered purity of \DevicePurity{} over 5~mm.  The criterion provides
physically constrained targets for subsequent full-vector design and
optimization.  It separates the generating material from
part of the dispersion-engineering task and provides a route for assessing
heterogeneous SPDC platforms that do not require periodic domain inversion.

\FloatBarrier
\section*{Appendices}

Appendices~\ref{sec:S_purity} and~\ref{sec:S_reachability} give the
dimensionless spectral model and the joint phase--delay derivation.
Appendices~\ref{sec:S_model}--\ref{sec:S_femwell} document the AlN/TiO$_2$
feasibility example, including the numerical methods, nonlinear overlap,
JSA, sensitivity, retuning, and independent solver check.
Appendices~\ref{sec:S_sic} and~\ref{sec:S_liquid} retain the alternative
solid and preliminary liquid examples; Appendix~\ref{sec:S_phasecheck}
examines the reduced-model consistency and its limitations.

All material-resolved results use constant-gap cross-sections.  The ideal
Gaussian-kernel calculation in Appendix~\ref{sec:S_gaussian} is a separate
analytical benchmark; no physical gap profile or near-unit-purity device is
reported.  The full-vector carrier, delay, and JSA results are distinct from
the splitting-calibrated two-mode consistency checks in
Appendix~\ref{sec:S_phasecheck}, which do not independently validate a
predictive screen based only on isolated fields.

Unless stated otherwise, displayed calculated results are rounded to three
decimal places, and mantissas in scientific notation use three significant
figures.  Integer design inputs and published dispersion coefficients retain
their native precision.  Cross-solver comparisons retain additional digits
where needed to resolve small differences; preliminary liquid and reduced-model
checks use fewer where appropriate.  All calculations use the unrounded
machine-readable data retained with the manuscript.

\appendix
\numberwithin{equation}{section}
\renewcommand{\theequation}{\thesection\arabic{equation}}

\section{Dimensionless purity model and numerical benchmarks}
\label{sec:S_purity}

Energy conservation fixes
\begin{equation}
 \omega_{p0}=\omega_{s0}+\omega_{i0}.
 \label{eq:S_energy}
\end{equation}
We parameterize the central colors by
\begin{equation}
 \eta=\frac{\omega_{s0}-\omega_{i0}}{\omega_{p0}},\qquad
 \omega_{s0}=\frac{1+\eta}{2}\omega_{p0},\qquad
 \omega_{i0}=\frac{1-\eta}{2}\omega_{p0}.
 \label{eq:S_eta}
\end{equation}
The reduced theory itself does not select 1300 or 1919~nm.  Absolute
wavelengths arise only after material dispersion and fabrication constraints
are included.  The device example has a pump at 775.000~nm, a signal at
1300.000~nm, and an idler at 1919.048~nm; these values
satisfy Eq.~\eqref{eq:S_energy} to the precision used by the mode solver.

Let $\nu_s=\omega_s-\omega_{s0}$ and
$\nu_i=\omega_i-\omega_{i0}$.  Throughout these appendices, a prime on a modal
propagation constant denotes its angular-frequency derivative at fixed
geometry and at the corresponding carrier:
$\beta'_\mu\equiv[\partial\beta_\mu(\omega)/\partial\omega]_{\omega=\omega_{\mu0}}$.
It is the group slowness, $1/v_{g,\mu}$, where $v_{g,\mu}$ is the modal group
velocity.  The same convention applies to isolated-guide and hybrid-mode
propagation constants.  Expansion of the exact stationary-mode mismatch gives
\begin{align}
 \Delta k(\nu_s,\nu_i)&=\Delta k_0+D_s\nu_s+D_i\nu_i
 +\frac{1}{2}G_{ss}\nu_s^2+G_{si}\nu_s\nu_i
 +\frac{1}{2}G_{ii}\nu_i^2+\cdots, \\
 D_s&=\beta'_p-\beta'_s,\qquad
 D_i=\beta'_p-\beta'_i.
 \label{eq:S_expansion}
\end{align}
The mode solver determines the delay ratio $\rho=D_i/D_s$; no delay value is
assigned by hand.  The physical optimization
minimizes both $|\Delta k_0|$ and $|D_s+D_i|$, targeting $\rho=-1$ when
$D_s\ne0$.

\subsection{Dimensionless constant-interaction benchmark}

Before introducing a material platform, we write the standard short-pulse JSA
in scale-free form~\cite{Grice1998,Law2000,URen2005}.  Here $\sigma_p$ is the
pump intensity-spectrum standard deviation in angular frequency,
\begin{equation}
 F_\rho(x,y)=\mathcal N\exp[-(x+y)^2/4]
 \operatorname{sinc}[B(x+\rho y)],
 \quad
 x=\frac{\nu_s}{\sigma_p},\quad y=\frac{\nu_i}{\sigma_p},\quad
 B=\frac{L|D_s|\sigma_p}{2}.
 \label{eq:S_dimensionlessJSA}
\end{equation}
Table~\ref{tab:S_uniform_benchmark} gives wide-window results, including two
bounded optimizations over $B$.  For
a uniform rectangular interaction, the symmetric target $\rho=-1$ approaches
$P=0.806$ at $B=1$.  The asymmetric-GVM limit $\rho=0$ reaches $P=0.955$
at $B=12$.  A physical design near this limit must retain higher-order
dispersion because the first-order idler coefficient vanishes.  Equal slopes,
$\rho=1$, do not give a factorable state.  Both factors in
Eq.~\eqref{eq:S_dimensionlessJSA} then depend only on $x+y$, and the apparent
finite-window purity decreases as the orthogonal window is enlarged.

\begin{table}[t]
 \centering
 \caption{Wide-window checks for the dimensionless constant-interaction model.  The
 half-span and grid spacing are in units of $\sigma_p$.  ``opt.'' marks a
 bounded optimization over $B$ with the same first-order model.  The
 $\pm9\sigma_p$ map and best-$B$ curve of
 Fig.~\ref{fig:puritymap}(a,b) give higher values, 0.833 at $\rho=-1$ and 0.960 at
 $\rho=0$, $B=12$.}
 \label{tab:S_uniform_benchmark}
 \begin{tabular}{@{}lrrrrr@{}}
  \toprule
  Regime & $\rho$ & $B$ & Half-span & Grid step & $P$\\
  \midrule
  Symmetric GVM & $-1$ & 1 & 60 & 0.100 & 0.806\\
  Best $B$ & $-2$ & opt. & 60 & 0.100 & 0.814\\
  Best $B$ & $-0.5$ & opt. & 60 & 0.100 & 0.814\\
  Asymmetric GVM & 0 & 12 & 40 & 0.050 & 0.955\\
  Equal slopes & 1 & 1 & 4 & 0.050 & 0.316\\
  Equal slopes & 1 & 1 & 40 & 0.050 & 0.033\\
  \bottomrule
 \end{tabular}
\end{table}

\subsection{Ideal Gaussian-kernel benchmark}
\label{sec:S_gaussian}

The symmetric constant-interaction result is limited by the sinc side lobes
of a rectangular longitudinal window.  Exact separability in the
linear-coordinate model requires Gaussian pump and phase-matching functions
~\cite{QuesadaBranczyk2018}.  Previous nonlinearity-shaping methods have
approximated Gaussian longitudinal kernels through domain engineering or
changes in propagation direction
~\cite{Branczyk2011,Dosseva2016,Tambasco2016,PovedaHospital2023}.  To keep the
derivation independent of an implementation, define an effective generation
kernel $g(z)$ after central phase matching.  With first-order slopes that are
constant along $z$,
\begin{equation}
 \Phi_g(\nu_s,\nu_i)=\int\!\rmd z\,g(z)
 \exp[\ii(D_s\nu_s+D_i\nu_i)z].
 \label{eq:S_generalkernel}
\end{equation}
For
\begin{equation}
 g_G(z)=g_0\exp\!\left[-\frac{(z-z_0)^2}{2\sigma_z^2}\right],
 \label{eq:S_gaussiankernel}
\end{equation}
the Fourier transform is, apart from a separable linear spectral phase,
\begin{equation}
 \Phi_G\propto\exp\!\left[-\frac{\sigma_z^2}{2}
 (D_s\nu_s+D_i\nu_i)^2\right].
 \label{eq:S_gaussianpm}
\end{equation}
Multiplication by the pump amplitude gives an exponent whose mixed term is
\begin{equation}
 -\left(\frac{1}{2\sigma_p^2}+\sigma_z^2D_sD_i\right)
 \nu_s\nu_i.
 \label{eq:S_gaussiancross}
\end{equation}
It vanishes for
\begin{equation}
 D_sD_i<0,\qquad
 \sigma_z^2=-\frac{1}{2\sigma_p^2D_sD_i}.
 \label{eq:S_gaussiancondition}
\end{equation}
The resulting first-order, untruncated JSA is exactly separable and has
$P=1$.  At $\rho=-1$, truncation at $\pm2\sigma_z$ gives $P=0.992$;
truncation at $\pm1.5\sigma_z$ gives 0.952.  These are normalized
Gaussian-kernel calculations, not material-resolved devices.  They explain
why the symmetric uniform-interaction benchmark near 0.81 is not a
fundamental purity limit of the coupled-waveguide mechanism.

For a smoothly varying separation $d(z)$, the corresponding lossless adiabatic
local-supermode approximation is
\begin{equation}
 \Phi(\omega_s,\omega_i)=\int_0^L\!\rmd z\,
 \Gamma(\omega_s,\omega_i;d(z))
 \exp\!\left[\ii\int_0^z\!\rmd z'\,
 \Delta k(\omega_s,\omega_i;d(z'))\right].
 \label{eq:S_varyinggap}
\end{equation}
Here $z'$ is a dummy longitudinal integration coordinate, not a derivative.
The local fields must be tracked with a consistent phase convention along the
selected branches, and interbranch conversion is neglected.  In this
formulation, $g(z)$ cannot in general be identified with the coupling
coefficient $\kappa(z)$.  A varying separation changes the local tensor
overlap, phase mismatch, delay slopes, participation, and coupling dispersion
simultaneously.  Realizing a prescribed generation kernel therefore requires
a separate inverse design, including a check of adiabaticity.  The present
finite-element database and Schmidt calculation apply only to the constant-gap
candidate.  They do not evaluate a Gaussian $g(z)$ or a $P>0.990$ gap profile.

\section{Joint phase--delay reachability derivation}
\label{sec:S_reachability}

In the isolated-guide basis, the selected idler subspace is represented by the
standard two-mode coupled-waveguide matrix
~\cite{Liu2019,Main2019PRA,Ding2021Heterogeneous,Shi2023},
\begin{equation}
 \mathbf H_i(\omega,d)=
 \begin{pmatrix}
  \beta_{iA}(\omega)&\kappa(\omega,d)\\
  \kappa(\omega,d)&\beta_{iB}(\omega)
 \end{pmatrix}.
 \label{eq:S_H}
\end{equation}
The local eigenvalues are
\begin{equation}
 \beta_{i,\pm}=\frac{\beta_{iA}+\beta_{iB}}{2}
 \pm\sqrt{\frac{(\beta_{iA}-\beta_{iB})^2}{4}+\kappa^2}.
 \label{eq:S_eigenvalues}
\end{equation}
The numerical database reports the half-splitting
$\Omega/2=(\beta_{i,+}-\beta_{i,-})/2$.  This equals the off-diagonal
coefficient in Eq.~\eqref{eq:S_H} only at exact isolated-guide resonance.
If $f_A^{(\mathrm{basis})}$ is obtained by projecting the supermode onto a
normalized isolated-guide basis, the two-mode identity is
\begin{equation}
 \kappa_{\mathrm{2m}}=\Omega
 \sqrt{f_A^{(\mathrm{basis})}[1-f_A^{(\mathrm{basis})}]}.
 \label{eq:S_kappa_inference}
\end{equation}
The full-vector device database instead reports the core-domain
electric-field-intensity share
\begin{equation}
 C_A=\frac{\int_A\rmd A\,\sum_a|E_a|^2}
 {\int_{A\cup B}\rmd A\,\sum_a|E_a|^2}.
 \label{eq:S_core_share}
\end{equation}
This diagnostic is useful for branch tracking and overlap screening; it is
neither a power fraction nor generally equal to
$f_A^{(\mathrm{basis})}$.  No isolated-mode projection was performed with the
COMSOL model, so Eq.~\eqref{eq:S_kappa_inference} is not applied to those
data.  Appendix~\ref{sec:S_phasecheck} carries out the projection with the
independent solver of Appendix~\ref{sec:S_femwell}, where it also compares
Eq.~\eqref{eq:S_kappa_inference} with a coupling inferred from the splitting
and the isolated detuning.  The raw half-splitting is
retained as a direct hybridization diagnostic, and all device predictions use
the calculated stationary-supermode propagation constants.

The nonlinear calculation uses one stationary eigenmode of the complete
cross-section, with central condition
\begin{equation}
 \beta_p(\omega_{p0})-\beta_s(\omega_{s0})
 -\beta_{i,\sigma}(\omega_{i0},d)=0,\qquad \sigma\in\{+,-\}.
 \label{eq:S_direct}
\end{equation}
Equation~\eqref{eq:S_direct} contains no reciprocal vector.  The nonlinear
coefficient is not periodically poled or sign reversed, and the constant-gap
device has no longitudinal modulation.  The process is direct modal phase
matching rather than quasi-phase matching (QPM).
The material-specific example below selects the lower-$\beta$ branch
($\sigma=-$).

\subsection{Phase closure fixes the participation}
\label{sec:S_phaseclosure}

The delay-only screen treats the participation as adjustable.  Joint
reachability additionally requires that this participation be the one selected
by carrier phase matching.  The relations below are the standard two-mode
eigenvalue and eigenvector results evaluated at that propagation constant.
Assume that the real symmetric
matrix in Eq.~\eqref{eq:S_H} acts on an orthonormal two-mode basis, take
$\kappa>0$ by a choice of basis phase, and keep the pump and signal in the
main guide, so that $D_s=\tau_s$, which we take to be nonzero.  At the fixed
carriers define
\begin{equation}
 \beta_*=\beta_{pA}-\beta_{sA},\qquad
 \Delta_A=\beta_*-\beta_{iA},\qquad
 \Delta_B=\beta_*-\beta_{iB}.
 \label{eq:S_phaseoffsets}
\end{equation}
Here $\Delta_A$ is the carrier phase mismatch of the isolated main guide.
Then
\begin{equation}
 \det(\beta_*\mathbf I-\mathbf H_i)=0
 \quad\Longleftrightarrow\quad
 \Delta_A\Delta_B=\kappa^2.
 \label{eq:S_phaseproduct}
\end{equation}
Both offsets must therefore have the same nonzero sign.  A target between the
two isolated propagation constants, or equal to either of them, cannot be
reached with nonzero real coupling at these inputs.  Changing the geometry or
the colors changes the inputs, so the condition does not exclude a material
platform.

For an eigenvector $(c_A,c_B)^{T}$ at $\beta_*$, the first row of
$\mathbf H_i$ gives $c_B/c_A=\Delta_A/\kappa\equiv u$.  Normalization,
Eq.~\eqref{eq:S_phaseproduct}, and the sign condition then give
\begin{equation}
 f_A^{\mathrm{PM}}=|c_A|^2=\frac{1}{1+u^2},\qquad
 \Delta_B=\frac{\kappa}{u},\qquad
 \sigma=\operatorname{sgn}(u).
 \label{eq:S_phasemixing}
\end{equation}
The normalized detuning is $t=(\beta_{iA}-\beta_{iB})/(2\kappa)=(1/u-u)/2$,
so this is the standard eigenvector parametrization written at the
phase-matched propagation constant, not a different coupled-mode mechanism.
The eigenvalue derivative at fixed geometry is the expectation value of
$\mathbf H_i'\equiv[\partial\mathbf H_i(\omega,d)/\partial\omega]_{\omega=\omega_{i0}}$
in the normalized eigenvector,
\begin{equation}
 \beta'_{i,\sigma}=
 \frac{\beta'_{iA}+u^2\beta'_{iB}+2u\kappa'}{1+u^2},
 \label{eq:S_phasegroup}
\end{equation}
where $\kappa'\equiv[\partial\kappa(\omega,d)/\partial\omega]_{\omega=\omega_{i0}}$
is the coupling dispersion.  All derivatives are evaluated at the idler
carrier with the separation and cross-section fixed, not along a re-optimized
phase-matching curve.  With
$r=\tau_i/\tau_s$, $m=(\beta'_{iB}-\beta'_{iA})/\tau_s$, and $q=\kappa'/\tau_s$,
\begin{equation}
 \rho_{\mathrm{PM}}=r-\frac{m u^2+2qu}{1+u^2}.
 \label{eq:S_phaseconditioned}
\end{equation}
Because $1-f_A^{\mathrm{PM}}=u^2/(1+u^2)$ and
$\sigma(1+t^2)^{-1/2}=2u/(1+u^2)$, Eq.~\eqref{eq:S_phaseconditioned} is the
main-text delay relation evaluated at the phase-matched point.  For a target $\rho_{\mathrm{tar}}$ and
$a=r-\rho_{\mathrm{tar}}$, the delay condition becomes
\begin{equation}
 (m-a)u^2+2qu-a=0.
 \label{eq:S_targetquadratic}
\end{equation}
A joint phase-and-delay solution needs a real, nonzero root that also
satisfies $u=\Delta_A/\kappa$ and Eq.~\eqref{eq:S_phaseproduct}; a root of
Eq.~\eqref{eq:S_targetquadratic} alone is therefore not a design.  For
$\tau_s>0$ and $m>0$, as in both designs of this work, $q<0$ means that the
coupling weakens with increasing frequency; the term $2qu$ then adds to the
correction $mu^2/(1+u^2)$ on the lower branch ($u<0$), the branch used in both
designs, which both have $q<0$ (Table~\ref{tab:S_phasecheck}).  If
$D_s=0$, the ratio coordinates must be replaced
by the unnormalized delays or the daughter labels interchanged.

For comparison, a frequency-independent QPM vector $K$ replaces $\beta_*$ by
$\beta_{pA}-\beta_{sA}-K$.  With periodic poling, $K$ absorbs whatever
mismatch remains, so the geometry can choose the detuning, and with it the
participation, freely.  Without poling, the detuning must itself close the
mismatch, and phase closure fixes the participation.  The two-mode algebra is
the same in both cases.

\FloatBarrier

\section{\texorpdfstring{AlN/TiO$_2$}{AlN/TiO2} full-vector numerical model}
\label{sec:S_model}

\subsection{Cross-section and material tensors}

The two-dimensional COMSOL Multiphysics 6.4 (Build 293) Electromagnetic
Waves, Frequency Domain model uses a propagation direction normal to the
transverse simulation plane.  The AlN $c$ axis is normal to the chip plane;
TiO$_2$ is modeled as isotropic.  Table~\ref{tab:S_geometry} summarizes the
final silica-embedded lateral geometry.
Top widths are quoted; the trapezoid bottom width is
\begin{equation}
 w_{\mathrm{bottom},j}=w_j+2h_j\cot\phi_j.
 \label{eq:S_bottomwidth}
\end{equation}
The two trapezoids share a baseline.  The quoted gap is the lateral distance
between their bottom edges and is constant along the propagation direction.

\begin{table}[h]
 \centering
 \caption{Nominal constant-gap candidate cross-section.}
 \label{tab:S_geometry}
 \begin{tabular}{lcc}
  \toprule
  Parameter & AlN main guide & TiO$_2$ auxiliary guide\\
  \midrule
  Film height (nm) & 1000 & 700\\
  Top width (nm) & 1091.388 & 481.161\\
  Sidewall angle & $80^\circ$ & $83^\circ$\\
  Bottom width (nm) & 1444.042 & 653.060\\
  Lateral bottom-edge gap (nm) & \multicolumn{2}{c}{450}\\
  Cladding & \multicolumn{2}{c}{dispersive silica}\\
  \bottomrule
 \end{tabular}
\end{table}

The relative-permittivity tensor is diagonal.  With wavelength in $\mu$m,
the implemented AlN ordinary and extraordinary indices are
~\cite{Bowman2018AlN}
\begin{align}
 n_{\mathrm{AlN},o}^2&=1+
 \frac{3.1425\lambda^2}{\lambda^2-0.153240^2},\\
 n_{\mathrm{AlN},e}^2&=1+
 \frac{3.3140\lambda^2}{\lambda^2-0.141710^2}.
 \label{eq:S_AlN}
\end{align}
The passive guide uses the provisional isotropic Cauchy function
\begin{equation}
 n_{\mathrm{TiO_2}}=2.39+\frac{0.0300}{\lambda^2}
 +\frac{0.00276}{\lambda^4}.
 \label{eq:S_TiO2}
\end{equation}
Silica is also dispersive.  Equation~\eqref{eq:S_TiO2} is a provisional model
and does not represent ellipsometry from a specified deposition.  We therefore
treat TiO$_2$ index variation explicitly below and do not call the nominal
dimensions a foundry-ready design.

\subsection{Hierarchical geometry search}

The search was organized around calculated eigenmode quantities; no
normalized delays were assigned by hand.  The AlN
height and angle were fixed at 1000~nm and $80^\circ$; the TiO$_2$ values were
fixed at 700~nm and $83^\circ$.  An isolated-guide search first identified
a narrow passive-guide mode near the idler phase index.  Coupled calculations
then tracked both idler branches while varying the two top widths and the
lateral gap.  Trials were rejected when the selected idler had negligible AlN
core share or when the pump and signal could not be assigned continuously
to the intended AlN modes.

For each trial, we solved for the AlN width by root-finding against
Eq.~\eqref{eq:S_direct}.  Group indices were then obtained from symmetric
frequency differences.  The theoretical purity map motivated a search near
$\rho=-1$.  Throughout the search, $\rho$ was calculated from the modal
dispersion.  The retained point uses a 450-nm gap, a 1091.388-nm AlN top width,
and a 481.161-nm TiO$_2$ top width.  It selects the lower-$\beta$ idler branch
and gives $\rho=-0.899$, rather than forcing exact symmetric matching.  This
is a locally selected feasibility point, not a global optimum over materials,
modes, colors, or cross-sections.

\subsection{Domain, mesh, branch tracking, and carrier metrics}

The transverse silica domain is $8~\mu$m wide and $6~\mu$m high.  The
maximum background element size is 150~nm.  The AlN core uses maximum/minimum
element sizes of 22/7~nm, and the TiO$_2$ core uses 14/5~nm.  These settings
define the production mesh used for the carrier search and broadband sweeps.
No independent refined broadband mesh and JSA were calculated for this
candidate, so the reported numbers should not be interpreted as an
asymptotic mesh-convergence bound.

Branch tracking does not assume that raw eigensolver indices remain fixed.
The pump is selected as an even higher-order TM-like mode by its dominant
$E_y$ polarization, AlN core share, parity, and transverse field moments.
The signal is the fundamental TM-like AlN mode.  The two idlers are selected
jointly by total AlN-plus-TiO$_2$ core share and ordered by propagation
constant.  Continuity of effective index, core-domain shares, and parity is
checked across frequency and geometry sweeps.  The chosen idler is the
lower-$\beta$ branch.  Its AlN share within the two cores is
\SIDeviceIdlerShare, while both pump and signal shares exceed 0.999.  The
eigensolver requests 38 pump modes and 28 daughter modes around branch-specific
effective-index shifts before applying these tests.

The unrounded carrier solutions give
\begin{align}
 \Delta k_0&=\SIDeviceMismatch~\mathrm{m^{-1}},\\
 (n_{\mathrm{eff},p},n_{\mathrm{eff},s},n_{\mathrm{eff},i})
 &=(\SIDeviceNeffP,\SIDeviceNeffS,\SIDeviceNeffI),\\
 (n_{g,p},n_{g,s},n_{g,i})
 &=(\SIDeviceNgP,\SIDeviceNgS,\SIDeviceNgI),\\
 (D_s,D_i)&=(\SIDeviceDs,\SIDeviceDi)~\mathrm{ps/m},\\
 \rho&=\SIDeviceRho,\qquad
 D_s+D_i=\SIDeviceSlope~\mathrm{ps/m}.
 \label{eq:S_nominalnumbers}
\end{align}
The 5-mm fixed-color phase-matching factor is 1.000 at the displayed
precision.  Group indices are
obtained from branch-tracked propagation constants according to
\begin{equation}
 n_g=c\frac{\partial\beta}{\partial\omega}.
 \label{eq:S_ng}
\end{equation}
The second-order dispersion values reconstructed at the carrier are
\begin{equation}
 (\beta_{2,p},\beta_{2,s},\beta_{2,i})
 =(-0.118,-0.157,-22.392)~\mathrm{ps^2/m}.
\end{equation}
The large idler curvature is why the device JSA retains exact branch
dispersion rather than relying only on the first-order delay ratio.

The broadband sweep samples the pump every 0.250~THz over $\pm2$~THz.  The
signal and both idler branches are sampled every 0.500~THz over $\pm12$~THz,
with 0.250-THz sampling over the central $\pm3$~THz interval.
Linear interpolation is used between calculated points; replacing it by
shape-preserving cubic or cubic-spline interpolation changes the purities below
by less than $10^{-4}$.  At $-5.0$ and $-4.5$~THz idler detuning, the tracked
lower-$\beta$ branch deviates from its smooth trend by $-1.9\times10^{2}$ and
$+3.5\times10^{2}~\mathrm{m^{-1}}$, visible as a small kink in main-text
Fig.~\ref{fig:devicevalidation}(c).  It is caused by a polarization anticrossing with a TE-like AlN mode
(Appendix~\ref{sec:S_femwell}); replacing the two affected samples by a spline
through their neighbors changes the purity by less than $10^{-5}$.  The
directly calculated idler half-splitting is
$\SIDeviceHalfSplit~\mathrm{m^{-1}}$.  Together with the core-domain shares,
it provides a direct hybridization diagnostic; no numerical two-mode coupling
coefficient is inferred from these COMSOL data (Appendix~\ref{sec:S_phasecheck}
estimates it with the independent solver).  The JSA uses the calculated stationary-supermode
propagation constant directly.

\section{Power normalization and full-vector nonlinear overlap}
\label{sec:S_overlap}

For each mode, forward power is calculated from
\begin{equation}
 P_\mu=\frac{1}{2}\operatorname{Re}\int
 \left(\mathbf E_\mu\times\mathbf H_\mu^*\right)\cdot\hat{\mathbf z}
 \,\rmd A.
 \label{eq:S_power}
\end{equation}
Only the AlN domain is included in the nonlinear integral.  In the crystal
coordinate system used by the model, the implemented complex numerator is
\begin{align}
 I_{33}={}&\int_{\mathrm{AlN}}d_{33}E_{p,y}E_{s,y}^*E_{i,y}^*\,\rmd A,\\
 I_{31}={}&\int_{\mathrm{AlN}}d_{31}E_{p,y}
 (E_{s,x}^*E_{i,x}^*+E_{s,z}^*E_{i,z}^*)\,\rmd A,\\
 I_{15}={}&\int_{\mathrm{AlN}}d_{15}\bigl[
 E_{p,x}(E_{s,x}^*E_{i,y}^*+E_{s,y}^*E_{i,x}^*)
 +E_{p,z}(E_{s,z}^*E_{i,y}^*+E_{s,y}^*E_{i,z}^*)\bigr]\,\rmd A.
 \label{eq:S_components}
\end{align}
The reported overlap is
\begin{equation}
 \Gamma=\frac{I_{33}+I_{31}+I_{15}}{\sqrt{P_pP_sP_i}},
 \label{eq:S_Gamma}
\end{equation}
with nominal magnitudes $|d_{33}|=4.3$~pm/V and
$|d_{31}|=|d_{33}|/45$~\cite{Majkic2017}.  The cited experiment reports
magnitudes rather than a relative sign and does not independently determine
$d_{15}$.  We use $|d_{15}|=|d_{31}|$ under Kleinman symmetry and take all
three coefficients to have a common sign in the nominal JSA.  The center
value is $|\Gamma|=\SIDeviceGamma~\mathrm{V^2/W^{3/2}}$.  Because $\Gamma$ is
complex, tensor contributions are added coherently before taking the
magnitude.  We evaluate Eq.~\eqref{eq:S_Gamma} directly; the AlN power
fraction is not used as an overlap proxy.  TiO$_2$ is assigned no quadratic
coefficient, so the AlN integral is the complete nonlinear channel represented
in this model.  The $d_{33}$ term contributes 99.989\% of the full-overlap
magnitude under the nominal tensor convention; the combined non-$d_{33}$
contribution is 0.011\%.  At the quoted precision, the unknown signs of the
two smaller terms have negligible effect on the overlap.
The tensor magnitudes are held constant across the modeled bandwidth.  Their
absolute wavelength dispersion and the quoted bulk-measurement uncertainty
are not propagated into a brightness error bar.  A common,
frequency-independent rescaling of the modeled AlN tensor cancels from the
normalized Schmidt purity; tensor dispersion need not cancel.

For the broadband overlap sweep, the pump and signal fields are held at
their central solutions while the selected idler supermode is recalculated
through the avoided crossing.  We approximate the numerical weighting as
$|\Gamma(\omega_s,\omega_i)|\simeq|\Gamma(\omega_{i})|$.  The pump and signal
propagation constants remain fully frequency dependent in
$\Delta k(\omega_s,\omega_i)$.  This approximation retains the rapid modal
change responsible for the overlap bandwidth while neglecting the slower
change of the two AlN-localized fields.  A constant-overlap control changes
the calculated purity from 0.789 to 0.797.  A calibrated brightness prediction
would still require a two-dimensional frequency sweep of the complex tensor
overlap, propagation loss, pump energy, and access-mode efficiencies.

\section{Constant-gap device JSA and Schmidt decomposition}
\label{sec:S_jsa}

For a transform-limited Gaussian pump with intensity full width at half
maximum $\Delta f_p$, the pump amplitude in ordinary-frequency detuning is
\begin{equation}
 \alpha(f_s+f_i)=
 \exp\left[-\frac{(f_s+f_i)^2}{4\sigma_I^2}\right],\qquad
 \sigma_I=\frac{\Delta f_p}{2\sqrt{2\ln2}}.
 \label{eq:S_pump}
\end{equation}
The calculated constant-gap amplitude is
\begin{align}
 F(f_s,f_i)={}&\mathcal N\,\Gamma(f_s,f_i)\alpha(f_s+f_i)
 \operatorname{sinc}\left[\frac{L\Delta k(f_s,f_i)}{2}\right]
 \exp\left[\frac{\ii L\Delta k(f_s,f_i)}{2}\right],\\
 \Delta k={}&\beta_p(\omega_{p0}+2\pi(f_s+f_i))
 -\beta_s(\omega_{s0}+2\pi f_s)
 -\beta_i(\omega_{i0}+2\pi f_i).
 \label{eq:S_exactJSA}
\end{align}
Here ``exact dispersion'' means that the sampled, branch-tracked propagation
constants are retained rather than replaced by a first-order expansion; it
does not imply numerically exact or mesh-converged modal solutions.
For $L=5$~mm, we optimize the pump bandwidth numerically and interpolate the
sampled idler-dependent vector-overlap magnitude across the avoided crossing.
The JSA is set to zero for $|f_s+f_i|>2$~THz, outside the calculated
pump-dispersion interval.  At the optimized bandwidth, this boundary lies far
into the Gaussian tail.
The optimized intensity FWHM is
\SIDevicePumpFWHM~THz, corresponding to a transform-limited intensity
duration of 1.176~ps.  Three calculations separate the model contributions:
\begin{align}
 P_{\mathrm{first\ order}}&=\SIDeviceFirstOrderPurity,\\
 P_{\mathrm{modal},\,|\Gamma|=\mathrm{const.}}&=\SIDeviceExactPurity,\\
 P_{\mathrm{modal},\,|\Gamma(f_i)|}&=\SIDevicePurity.
 \label{eq:S_puritycontrols}
\end{align}
All use the same pump bandwidth and a $\pm\SIDeviceHalfWindow$-THz grid.  The
change from first-order to modal dispersion reflects higher-order
stationary-mode dispersion.  The subsequent change shows the additional
effect of overlap variation.  The outer grid cells of the final JSA contain
probability $\SIDeviceEdgeProbability$.

For the constant-overlap control, widening the half-window from 6 to 12~THz
changes the calculated purity as shown in Table~\ref{tab:S_window}.  The trend
comes from accumulating weak sinc side lobes.  The final overlap-weighted
value is therefore reported explicitly as a 12-THz calculated-window result;
we do not extrapolate it to infinite bandwidth.

\begin{table}[t]
 \centering
 \caption{Finite-window check for the exact-dispersion, constant-overlap
 control at 0.080-THz grid spacing.}
 \label{tab:S_window}
 \begin{tabular}{@{}rrr@{}}
  \toprule
  Half-window (THz) & Purity & Edge probability\\
  \midrule
  6 & 0.804 & $2.19\times10^{-4}$\\
  8 & 0.801 & $2.28\times10^{-4}$\\
  10 & 0.799 & $1.27\times10^{-4}$\\
  12 & 0.797 & $1.64\times10^{-4}$\\
  \bottomrule
 \end{tabular}
\end{table}

On a uniform rectangular frequency grid, the Schmidt decomposition
~\cite{Law2000,Eberly2006} gives
\begin{equation}
 F_{jk}=\sum_n s_n u_{n,j}v_{n,k},\qquad
 \lambda_n=\frac{s_n^2}{\sum_m s_m^2},\qquad
 P=\sum_n\lambda_n^2,\qquad K=P^{-1}.
 \label{eq:S_Schmidt}
\end{equation}
Because the frequency spacings are equal and constant, their quadrature
factors cancel in the normalized $\lambda_n$.  The final 0.040-THz grid gives
a leading Schmidt weight of 0.888 and
$K=\SIDeviceSchmidt$.  No spectral filter is included.  Reporting the window
and edge probability identifies the result as a finite-window calculation.
This avoids interpreting a restricted numerical range as an intrinsic purity,
consistent with the finite-range caution in Ref.~\cite{Graffitti2018}.

\section{Sensitivity and retuning}
\label{sec:S_sensitivity}

\subsection{Fixed-color geometry perturbations}

Each geometric parameter was changed independently while the wavelengths and
all other dimensions were held at their nominal values.  Table
~\ref{tab:S_geometry_tolerance} reports the resulting central mismatch,
5-mm phase-matching factor, and idler AlN share.  These perturbations are
diagnostics, not measured process distributions or a Monte Carlo yield model.

\begin{table*}[t]
 \centering
 \caption{One-at-a-time fixed-color geometry sensitivity.  Width, height,
 and gap changes are in nm; angle changes are in degrees.}
 \label{tab:S_geometry_tolerance}
 \begin{tabular}{lrrr}
  \toprule
  Perturbation & $\Delta k_0$ (m$^{-1}$) &
  $\operatorname{sinc}^2(\Delta k_0L/2)$ & Idler AlN share\\
  \midrule
  Nominal & 4.710 & 1.000 & 0.632\\
  Gap $-30$ & 2742.717 & 0.006 & 0.628\\
  Gap $+30$ & $-2435.233$ & 0.001 & 0.636\\
  AlN width $-20$ & $-27112.149$ & $2.06\times10^{-4}$ & 0.692\\
  AlN width $+20$ & 26450.247 & $5.25\times10^{-6}$ & 0.565\\
  TiO$_2$ width $-20$ & 23784.374 & $1.46\times10^{-5}$ & 0.284\\
  TiO$_2$ width $+20$ & $-10101.420$ & $2.28\times10^{-5}$ & 0.873\\
  AlN height $-20$ & 9532.445 & 0.002 & 0.779\\
  AlN height $+20$ & $-5722.658$ & 0.005 & 0.445\\
  TiO$_2$ height $-20$ & 33292.650 & $1.44\times10^{-4}$ & 0.365\\
  TiO$_2$ height $+20$ & $-11396.755$ & $5.74\times10^{-5}$ & 0.905\\
  AlN angle $-1$ & 23429.995 & $2.35\times10^{-4}$ & 0.580\\
  AlN angle $+1$ & $-24196.655$ & $1.41\times10^{-4}$ & 0.677\\
  TiO$_2$ angle $-1$ & $-7458.206$ & $1.18\times10^{-4}$ & 0.803\\
  TiO$_2$ angle $+1$ & 12380.157 & $2.10\times10^{-4}$ & 0.418\\
  \bottomrule
 \end{tabular}
\end{table*}

Each geometry change affects two coupled quantities: it moves the carrier
mismatch and redistributes the hybrid idler between the two cores.
Consequently, restoring only $\Delta k_0=0$ need not restore the group-delay
ratio or nonlinear overlap.  The nominal dimensions do not support a
fixed-color fabrication-robust claim.

\subsection{\texorpdfstring{TiO$_2$}{TiO2} index uncertainty and two-width
correction}

We next add a wavelength-independent offset $\delta n$ to
Eq.~\eqref{eq:S_TiO2}.  At fixed geometry, offsets from $-0.020$ to $+0.020$
move $\Delta k_0$ from $40251.414$ to $-12440.290~\mathrm{m^{-1}}$ and the
idler AlN share from 0.200 to 0.928.  This range is a model-sensitivity scan,
not an estimate of actual TiO$_2$ film variation.

A targeted two-width calculation was then used for
$\delta n=-0.005,0,+0.005$.  At each offset, the AlN and TiO$_2$ top widths
were adjusted to recover central phase matching and a delay ratio near
$-0.9$.  Table~\ref{tab:S_index_correction} shows the selected solutions.

\begin{table*}[t]
 \centering
 \caption{Two-width design correction for the provisional TiO$_2$ index.
 The exact broadband JSA was calculated only for the nominal row.  The
 $\delta n=0$ row is the nominal geometry re-solved in the retuning run; its
 0.002-nm width difference from Table~\ref{tab:S_geometry} accounts for the
 2-m$^{-1}$ difference in $\Delta k_0$.}
 \label{tab:S_index_correction}
 \begin{tabular}{rrrrrrr}
  \toprule
  $\delta n$ & $w_{\mathrm{AlN}}$ (nm) & $w_{\mathrm{TiO_2}}$ (nm) &
  $\Delta k_0$ (m$^{-1}$) & $\rho$ & $\Omega/2$ (m$^{-1}$) & AlN share\\
  \midrule
  $-0.005$ & 1091.307 & 488.812 & 0.841 & $-0.902$ & 22294.072 & 0.624\\
  0 & 1091.386 & 481.161 & 2.617 & $-0.899$ & 22534.109 & 0.632\\
  $+0.005$ & 1091.396 & 473.565 & 0.529 & $-0.902$ & 22768.533 & 0.637\\
  \bottomrule
 \end{tabular}
\end{table*}

\begin{figure*}[t]
 \centering
 \includegraphics[width=0.95\textwidth]
 {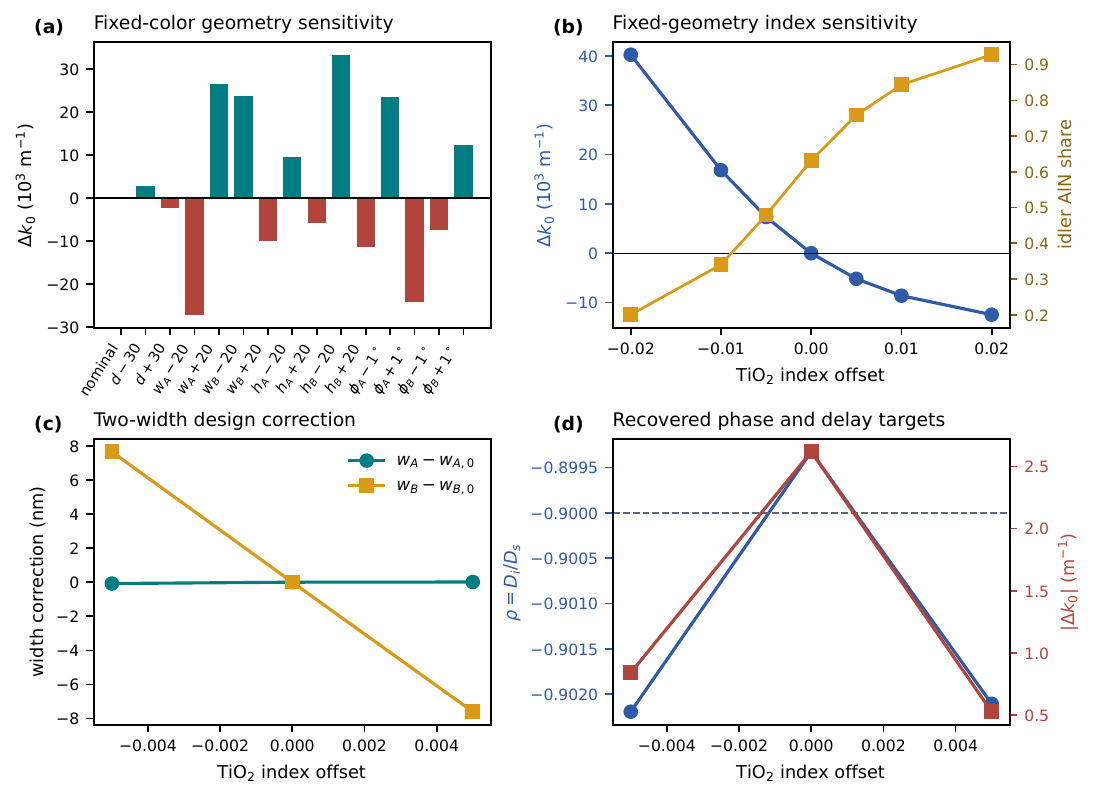}
 \caption{Sensitivity and design-stage correction of the AlN/TiO$_2$
 candidate.  (a) Fixed-color mismatch under one-at-a-time geometry changes.
 (b) Fixed-geometry response to a TiO$_2$ index offset.  (c) Width corrections
 selected at $\delta n=-0.005,0,+0.005$.  (d) Recovered central mismatch and
 delay ratio.}
 \label{fig:S_tolerance_controls}
\end{figure*}

Figure~\ref{fig:S_tolerance_controls} collects the fixed-geometry sensitivity
and the two-width design correction.

Figure~\ref{fig:S_tolerance_controls}(a) shows that the width, height, and
sidewall-angle perturbations move the fixed-color mismatch by substantially
more than the gap perturbation, although every tested perturbation reduces the
5-mm phase-matching factor below 0.007.  In
Fig.~\ref{fig:S_tolerance_controls}(b), increasing the TiO$_2$ index offset
moves the mismatch monotonically through zero while transferring the idler
field from TiO$_2$ toward AlN.  The two-width correction in
Fig.~\ref{fig:S_tolerance_controls}(c) is carried mainly by the TiO$_2$ width,
which changes by approximately $\pm7.6$~nm at $\delta n=\mp0.005$; the AlN
width changes by less than 0.1~nm.  These paired adjustments recover
$|\Delta k_0|<0.9~\mathrm{m^{-1}}$ and $\rho\simeq-0.902$ at both endpoints,
as shown in Fig.~\ref{fig:S_tolerance_controls}(d).  Two independent geometric
controls can therefore restore the carrier and first-order delay targets after
the passive-film index is measured.

\subsection{Fixed-pump retuning after a width error}
\label{sec:S_retune}

At fixed colors, a 1-nm change of the AlN top width moves $\Delta k_0$ by about
$1.3\times10^{3}~\mathrm{m^{-1}}$ (Table~\ref{tab:S_geometry_tolerance}), so a
5-mm device is phase matched at the nominal colors only within a fraction of a
nanometer.  In SPDC, however, a fixed pump generates the daughter pair at
whatever colors satisfy phase matching.  We therefore kept the pump at 775~nm
in the femwell model and solved for the signal frequency detuning $f_s$ (idler
detuning $-f_s$) at which the lower-branch mismatch vanishes, first for the
nominal width and then for AlN widths changed by $\pm1$~nm.  Because femwell
gives $\Delta k_0=135~\mathrm{m^{-1}}$ at the nominal colors, its own
nominal phase-matched point lies at $f_s=-0.027$~THz.
Table~\ref{tab:S_retune} lists the results.  Relative to that point, the
phase-matched signal moves by $+0.26$ and $-0.26$~THz, well
within the calculated branch data, and the delay ratio changes from
$-0.903$ to $-0.975$ and $-0.830$ because the idler moves
along the strongly dispersive hybrid branch.  For errors of about a nanometer,
a width error therefore shifts the source colors and the delay ratio; it does
not remove phase matching.  Larger errors move the idler further along the
branch, where $\rho$ changes by roughly 0.07 per nanometer.

\begin{table}[h]
 \centering
 \caption{Fixed-pump retuning of the AlN/TiO$_2$ design after a $\pm1$-nm AlN
 width change (femwell).  $f_s$ is the signal frequency detuning of the
 phase-matched point from 1300~nm.}
 \label{tab:S_retune}
 \begin{tabular}{rrrrr}
  \toprule
  $\delta w_{\mathrm{AlN}}$ (nm) & $f_s$ (THz) & $D_s$ (ps/m) & $D_i$ (ps/m) & $\rho$\\
  \midrule
  $-1$ & $+0.236$ & 427.8 & $-417.3$ & $-0.975$\\
  0 & $-0.027$ & 427.1 & $-385.7$ & $-0.903$\\
  $+1$ & $-0.289$ & 426.4 & $-353.8$ & $-0.830$\\
  \bottomrule
 \end{tabular}
\end{table}

\section{Independent mode-solver check of the AlN/TiO\texorpdfstring{$_2$}{2} point}
\label{sec:S_femwell}

To test whether the carrier solution depends on the COMSOL discretization, we
recomputed the nominal cross-section with an independent finite-element solver
(femwell~\cite{femwell} on scikit-fem~\cite{skfem}), using second-order
N\'ed\'elec transverse and Lagrange longitudinal elements on a conforming mesh
whose element edges follow every trapezoid side, Malitson
silica~\cite{Malitson1965}, the AlN and TiO$_2$ models of
Eqs.~\eqref{eq:S_AlN} and \eqref{eq:S_TiO2}, and a perfectly conducting
boundary on the same $8\times6~\mu$m domain.  Group indices were obtained with
the same five-point stencil ($h=0.25$~THz) as the COMSOL data.
Table~\ref{tab:S_femwell} compares the carrier quantities.  The effective
indices agree to within $6\times10^{-5}$ (pump and signal to
$10^{-6}$), the central mismatch is
$135~\mathrm{m^{-1}}$, equivalent to a
$0.1$-nm change of the AlN width, and $\rho=-0.912$ at the nominal
colors.  At the femwell model's own phase-matched point, $-0.027$~THz from
the nominal signal frequency, $\rho=-0.903$
(Appendix~\ref{sec:S_retune}).  Refining the
core elements from 20 to 14~nm changes $n_{\mathrm{eff}}$ by at most
$2\times10^{-6}$ and $\Delta k_0$ by $6~\mathrm{m^{-1}}$.  The remaining
difference in $\rho$ ($-0.903$ versus $-0.899$ at the respective
phase-matched points) reflects the steep dependence of the idler group index on
the detuning near the avoided crossing; the idler share (0.628 versus 0.632)
shows a slightly different detuning in the two models.  The same solver also
shows that the TM-like lower idler branch crosses a TE-like AlN mode near
$-4.6$~THz idler detuning: between $-4.75$ and $-4.5$~THz the two modes
exchange polarization character ($E_y$ fraction about 0.2 and 0.6) at an
effective-index separation of about $6\times10^{-4}$.

\begin{table}[h]
 \centering
 \caption{Nominal AlN/TiO$_2$ carrier quantities from COMSOL and from the
 independent femwell calculation (20-nm core elements).}
 \label{tab:S_femwell}
 \begin{tabular}{lrr}
  \toprule
  Quantity & COMSOL & femwell\\
  \midrule
  $n_{\mathrm{eff},p}$ & 1.90867 & 1.90867\\
  $n_{\mathrm{eff},s}$ & 1.96001 & 1.96001\\
  $n_{\mathrm{eff},i}$ (lower branch) & 1.83288 & 1.83284\\
  $n_{g,p},\,n_{g,s},\,n_{g,i}$ & 2.3273, 2.1992, 2.4425 & 2.3273, 2.1992, 2.4440\\
  $\Delta k_0$ (m$^{-1}$) & 4.7 & 135\\
  $(D_s,D_i)$ (ps/m) & $(427.1,-384.1)$ & $(427.1,-389.4)$\\
  $\rho$ & $-0.899$ & $-0.912$\\
  $\beta_{2,i}$ (ps$^2$/m) & $-22.39$ & $-22.44$\\
  Idler AlN share & 0.632 & 0.628\\
  \bottomrule
 \end{tabular}
\end{table}

\section{Alternative auxiliary material: AlN/4H-SiC}
\label{sec:S_sic}

To test whether the coupled-supermode mechanism depends on TiO$_2$ or on a
strictly linear auxiliary, we also analyze a separately optimized
AlN/SiO$_2$/4H-SiC calculation from the material screen.  This comparison uses
a vertical rather than lateral coupler and was not used to select or tune the
AlN/TiO$_2$ point.  It tests material replaceability under the same
requirements of direct phase matching, opposite-sign daughter delays, and
nonzero field in the AlN generation region.

The full-vector COMSOL model contains a $1000$-nm-high AlN lower guide with a
$1183.551$-nm top width and an $80^\circ$ sidewall angle.  A $150$-nm uniform
silica spacer separates it from a centered, $700$-nm-high 4H-SiC upper guide
with a $250$-nm top width and an $83^\circ$ sidewall angle.  The pump, signal,
and idler wavelengths are $775.000$, $1343.084$, and $1832.282$~nm.  The pump
and signal are AlN-localized, and the selected idler is the higher-$\beta$
member of the hybrid pair.  AlN dispersion follows Eq.~\eqref{eq:S_AlN}; the
uniaxial 4H-SiC Sellmeier model is taken from Ref.~\cite{Wang2013SiC}.

\begin{figure*}[t]
 \centering
 \includegraphics[width=0.98\textwidth]
 {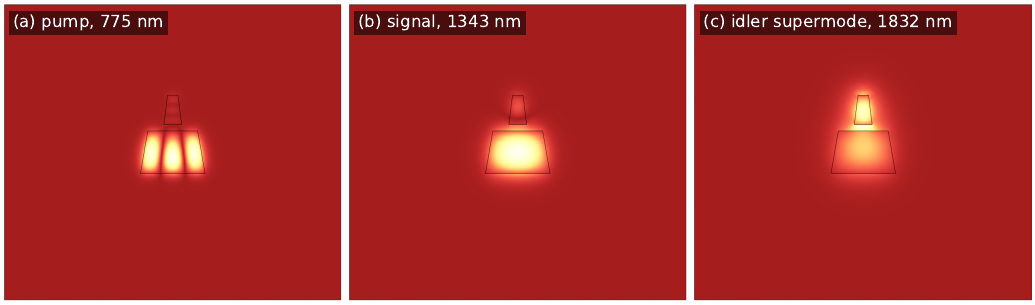}
 \caption{Full-vector mode profiles of the alternative vertical
 AlN/SiO$_2$/4H-SiC calculation.  Pump and signal remain predominantly in
 AlN, whereas the selected idler hybridizes across the two cores.  Each panel
 is normalized independently and shows the magnitude of the dominant field
 component.}
 \label{fig:S_sic_modes}
\end{figure*}

Figure~\ref{fig:S_sic_modes} shows the three selected mode profiles in this
vertical cross-section.

The mode maps in Fig.~\ref{fig:S_sic_modes} resolve the selective
hybridization used in the alternative design.  The higher-order pump in
Fig.~\ref{fig:S_sic_modes}(a) and the fundamental signal in
Fig.~\ref{fig:S_sic_modes}(b) remain concentrated in the lower AlN core.  By
contrast, the idler in Fig.~\ref{fig:S_sic_modes}(c) has appreciable field in
both the AlN and upper 4H-SiC guides.  This spatial distribution produces the
calculated idler AlN share of 0.589 while retaining the AlN overlap needed for
pair generation.

Table~\ref{tab:S_sic_metrics} summarizes the carrier and spectral checks.
The production-mesh point gives a nearly vanishing central mismatch and
$\rho\simeq-1$.  Re-evaluation on a refined mesh changes the fixed-geometry
mismatch to $77.789~\mathrm{m^{-1}}$ and the delay ratio to $-1.005$; the
corresponding 5-mm phase-matching factor remains $0.987$.  The largest relative
change among the three central effective indices is $1.00\times10^{-5}$.

\begin{table*}[t]
 \centering
 \caption{Alternative vertical AlN/4H-SiC calculation.  The JSA quantities
 refer only to the AlN generation integral.}
 \label{tab:S_sic_metrics}
 \small
 \begin{tabular}{lr}
  \toprule
  Quantity & Value\\
  \midrule
  $\Delta k_0$ (production mesh) & $-0.335~\mathrm{m^{-1}}$\\
  $\Delta k_0$ (refined mesh) & $77.789~\mathrm{m^{-1}}$\\
  $(n_{g,p},n_{g,s},n_{g,i})$ & $(2.326,2.216,2.435)$\\
  $(D_s,D_i)$ & $(364.350,-364.286)~\mathrm{ps/m}$\\
  $\rho=D_i/D_s$ (production/refined) & $-1.000/-1.005$\\
  Idler AlN share & 0.589\\
  $|\Gamma_{\mathrm{AlN}}|$ & $2.288\times10^{-3}~\mathrm{V^2/W^{3/2}}$\\
  Pump intensity FWHM & $0.416~\mathrm{THz}$\\
  Calculated-window purity / Schmidt number & $0.804/1.243$\\
  Inverse-window purity range & 0.798--0.800\\
  Edge probability & $8.00\times10^{-5}$\\
  \bottomrule
 \end{tabular}
\end{table*}

The 5-mm JSA uses branch-tracked COMSOL propagation constants over a
$\pm12$-THz half-window and the sampled idler dependence of the
power-normalized AlN vector overlap.  The optimized calculated-window result
is $P=0.804$; holding the overlap constant gives $P=0.805$, and first-order
dispersion gives $P=0.807$.  Increasing the frequency-grid resolution from
$0.040$ to $0.030$~THz changes the purity by $3.36\times10^{-6}$.  The
remaining difference from unity is again dominated by the uniform
rectangular-interaction sinc structure.

\begin{figure*}[t]
 \centering
 \includegraphics[width=0.98\textwidth]
 {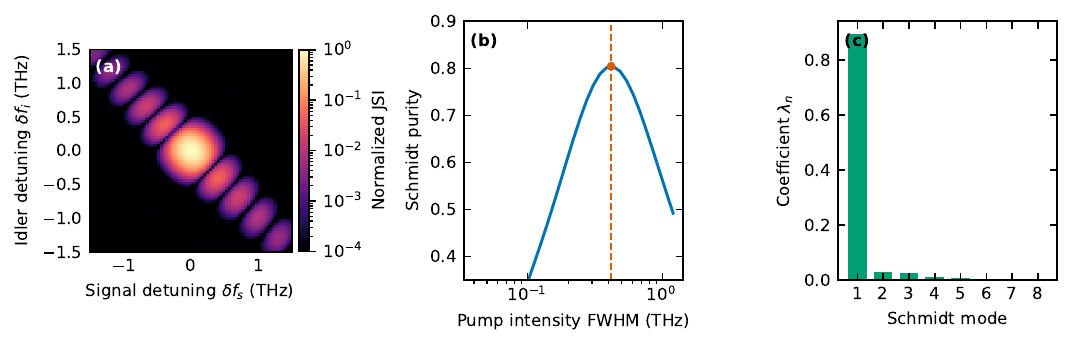}
 \caption{Spectral calculation for the AlN-generated channel of the
 alternative AlN/4H-SiC design.  The panels show the joint spectral intensity,
 pump-bandwidth optimization, and leading Schmidt weights.  No spectral
 filter is included.}
 \label{fig:S_sic_jsa}
\end{figure*}

Figure~\ref{fig:S_sic_jsa} summarizes the corresponding spectral calculation
and Schmidt decomposition.

Figure~\ref{fig:S_sic_jsa}(a) displays a compact central joint spectrum with
the diagonal sinc sidelobes of the uniform interaction.  The pump-bandwidth
scan in Fig.~\ref{fig:S_sic_jsa}(b) has a clear maximum at 0.416~THz, where the
calculated-window purity reaches 0.804.  The Schmidt spectrum in
Fig.~\ref{fig:S_sic_jsa}(c) is correspondingly dominated by its first mode,
with much smaller higher-order weights producing $K=1.243$.  The mode profiles,
phase-matching metrics, and spectral decomposition therefore give the same
near-symmetric constant-interaction signature as the lateral AlN/TiO$_2$
example in a different material stack.

Crystalline 4H-SiC is noncentrosymmetric and can itself contribute to SPDC.
The reported $P=0.804$ therefore characterizes the AlN-generated channel; a
coherent dual-material JSA would additionally require the SiC tensor, crystal
orientation, and relative phase.  The SiC result is used as a
material-replaceability check on the supermode mechanism.

Existing 4H-SiC-on-AlN photonics provides a fabrication precedent for the
material stack~\cite{Wang2024SiCAlN}.  It does not validate the exact vertical
cross-section modeled here.  One-at-a-time perturbations of $\pm10$~nm in
widths, heights, and spacer and $\pm1^\circ$ in sidewall angles show substantial
fixed-color sensitivity.
A $25$-nm lateral overlay retains a 5-mm phase-matching factor of 0.984 and
$\rho=-1.007$, whereas exact fixed-pump daughter-color retuning restores both
phase and $|\rho+1|\le0.2$ in only three of eight more sensitive cases.  These
results indicate that a process matrix and post-fabrication selection would be
needed.

\FloatBarrier
\section{Preliminary liquid-generation example: a ferroelectric nematic channel}
\label{sec:S_liquid}

\emph{The broadband dispersion, vector overlap, and JSA in this section use
the femwell model of Appendix~\ref{sec:S_femwell}.  An independent COMSOL 6.4
calculation checks the carrier modes and local group delays, but not the
quoted overlap or purity.}

\paragraph{Model.}  The nonlinear guide is a rectangular channel,
$2.63\times1.6~\mu$m, filled with the room-temperature ferroelectric
nematic liquid crystal FNLC-1751 and embedded in silica.  The polar (director)
axis lies along $x$, in the chip plane and normal to propagation, as can be set
by surface alignment or an in-plane field.  The liquid is modeled as uniaxial,
$\bm\varepsilon=\mathrm{diag}(n_e^2,n_o^2,n_o^2)$, with the Cauchy dispersion
reported for this material~\cite{Sultanov2024LiquidSPDC},
$n_e=1.663+10256~\mathrm{nm^2}/\lambda^2$ and
$n_o=1.474+874~\mathrm{nm^2}/\lambda^2$.  The auxiliary guide is a
$1.279\times0.4~\mu$m Si$_3$N$_4$ wire (dispersion from
Ref.~\cite{Luke2015SiN}), centred at the mid-height of the channel with a
0.5-$\mu$m edge-to-edge gap.  The colors are those of the AlN/TiO$_2$ design,
775, 1300, and 1919~nm.  All three fields are TE-like ($E_x$ dominant), so the
dominant tensor element is $d_{33}$ along the polar axis.  The liquid core is
FNLC-1751 throughout; only $d_{33}$ is taken from a different mixture.  We assume
$d_{33}=11$~pm/V as a proxy.  This value was measured for a different
ferroelectric nematic mixture, FNLC-1571, near $23^\circ$C by
second-harmonic generation driven at 800~nm~\cite{Lovsin2026FNLC}; it is not
a measurement of FNLC-1751 at the three wavelengths considered here.
For FNLC-1751 itself, only an estimate of about 20~pm/V is
available~\cite{Sultanov2024LiquidSPDC}.  The overlap values below scale
linearly with this assumed $d_{33}$.  The femwell mesh uses 50-nm elements in
the liquid and 20-nm elements in Si$_3$N$_4$; refining them to 35 and 14~nm changes
$\Delta k_0$ by 1.8~m$^{-1}$ and $\rho$ by 1$\times10^{-4}$.
We set $x=0$ at the centre of the gap and $y=0$ at the bottom of the
liquid channel.  At the nominal dimensions, the femwell domain extends
from $x=-6.88$ to $5.53~\mu$m and from $y=-4.0$ to $5.6~\mu$m,
leaving 4~$\mu$m of silica beyond the outermost core edges on each side.

\paragraph{Design.}  The isolated channel phase matches an even, higher-order
TE pump mode with the fundamental TE signal and idler near a width of
2.70~$\mu$m.  There, however, the isolated-channel delays
$\tau_\mu=\beta'_p-\beta'_\mu$ have the same sign and nearly equal size,
$r=\tau_i/\tau_s\simeq1.02$, which is close to the unfactorable equal-slope
case of Fig.~\ref{fig:puritymap}(c), so an auxiliary guide is needed to reverse the
idler delay.  A main guide that is phase matched on its own cannot stay phase
matched at the same colors once the auxiliary is coupled
(Eq.~\eqref{eq:S_phaseproduct}), so the design used the order of
Appendix~\ref{sec:S_phaseclosure}.  Narrowing the channel to $2.63~\mu$m sets
its carrier mismatch to $\Delta_A=-1.46\times10^{4}~\mathrm{m^{-1}}$, about
0.71 times the idler coupling.  The negative sign selects the lower
branch ($u<0$), with $f_A^{\mathrm{PM}}\simeq0.66$
(Appendix~\ref{sec:S_phasecheck}).  The Si$_3$N$_4$ width was then tuned until
the femwell lower idler supermode closed the carrier mismatch.  In the
two-mode model this is the condition $\Delta_B=\kappa^2/\Delta_A$, which the
isolated-mode values meet to within about 22\%
(Appendix~\ref{sec:S_phasecheck}).  The Si$_3$N$_4$ fraction
raises the idler group index from 1.706 (isolated channel) to 1.821 and
reverses the sign of $D_i$.  A larger detuning, $\Delta_A=-2.63\times10^{4}~\mathrm{m^{-1}}$
at a channel width of 2.58~$\mu$m, followed by the same Si$_3$N$_4$ retuning,
gives a smaller liquid share (core-domain intensity share 0.56) and
$\rho=-1.74$.  The channel width therefore tunes $\rho$, in the
direction predicted by Eq.~\eqref{eq:S_phaseconditioned}.  The design thus
changed both guides: the channel width (from 2.70 to 2.63~$\mu$m) and the
Si$_3$N$_4$ width, not the auxiliary guide alone.

\begin{figure*}[t]
 \centering
 \includegraphics[width=0.98\textwidth]{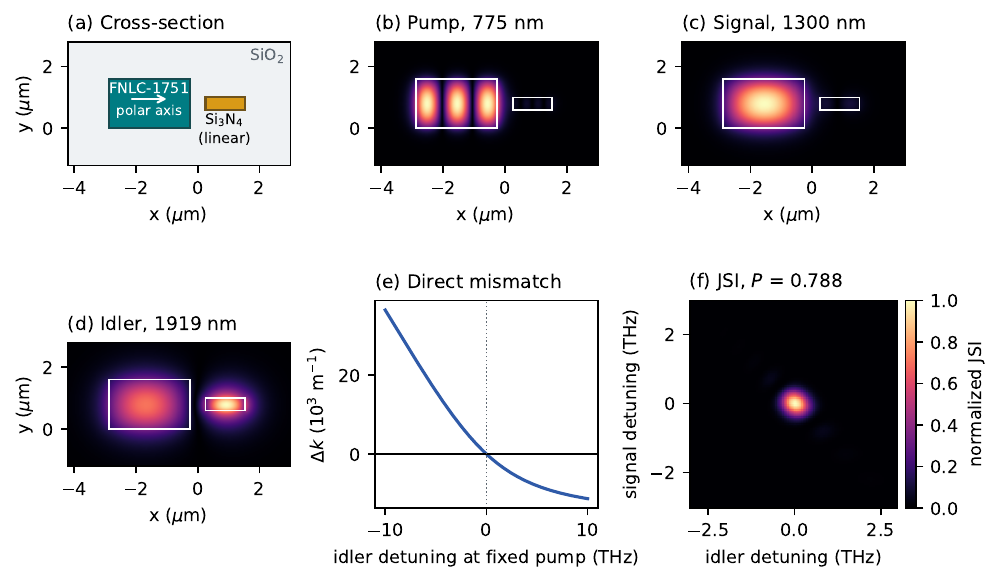}
 \caption{Preliminary liquid-generation example.  (a) Cross-section:
 FNLC-1751 channel (polar axis along $x$) and Si$_3$N$_4$ wire in silica.
 (b)--(d) $|E_x|$ of the higher-order pump, the signal, and the lower idler
 supermode.  (e) Direct mismatch along the fixed-pump energy-conservation line.
 (f) Joint spectral intensity at the purity-optimized pump bandwidth.}
 \label{fig:S_liquid}
\end{figure*}

\paragraph{Results.}  Table~\ref{tab:S_liquid} and
Fig.~\ref{fig:S_liquid} summarize the calculation.  With a Gaussian pump
optimized for purity, the exact-dispersion JSA over 5~mm with a constant
overlap gives $P=0.788$ (first-order dispersion: 0.808) at a
pump intensity FWHM of 0.813~THz, over a $\pm18$-THz window with
edge probability 1.4$\times10^{-4}$.  The window stops below $+19$~THz signal
detuning, where the signal starts to hybridize with a Si$_3$N$_4$ mode.  As for
AlN/TiO$_2$, the purity still decreases slowly with the window (0.806, 0.797,
0.791, and 0.788 at $\pm8$, 12, 16, and 18~THz) because of the sinc side
lobes; the like-for-like AlN/TiO$_2$ value, with constant overlap over
$\pm12$~THz, is 0.797.  The geometric part of the central
power-normalized overlap is 0.91 times that of the AlN/TiO$_2$ design,
evaluated with the same solver and normalization.  With the assumed proxy
$d_{33}=11$~pm/V the overlap is therefore 2.3 times that of AlN
($d_{33}=4.3$~pm/V), and 4.2 times with the 20-pm/V estimate for
FNLC-1751.  The same calculation reproduces the COMSOL AlN/TiO$_2$ value,
$|\Gamma|=2.77\times10^{-3}~\mathrm{V^2/W^{3/2}}$ ($d_{33}$ term only), so the
liquid value for the assumed 11~pm/V is
$|\Gamma|\simeq6.4\times10^{-3}~\mathrm{V^2/W^{3/2}}$.  The remaining nonlinear-tensor elements,
propagation loss, scattering from director fluctuations, alignment disorder,
and the validity of the Cauchy law near 1.9~$\mu$m are not included.  The
example therefore provides preliminary numerical evidence for direct phase
matching with opposite-sign pump--daughter delays in a liquid generation region.

\begin{table}[h]
 \centering
 \caption{Carrier quantities of the preliminary FNLC/Si$_3$N$_4$ example at
 the nominal dimensions, from femwell and from the independent COMSOL carrier
 check.  COMSOL checks the carrier modes, local group delays, and idler field
 share; the remaining rows are femwell results (--, not computed).
 The field share is the liquid-core integral of $|\mathbf{E}_i|^2$ divided
 by its integral over both cores, excluding the cladding.}
 \label{tab:S_liquid}
 \small
 \begin{tabular}{lrr}
  \toprule
  Quantity & femwell & COMSOL 6.4\\
  \midrule
  Channel / Si$_3$N$_4$ width ($\mu$m) & 2.63 / 1.279 & 2.63 / 1.279\\
  $\Delta k_0$ (m$^{-1}$) & 14.4 & $-95.5$\\
  $n_{\mathrm{eff},p}$ & 1.605866 & 1.605860\\
  $n_{\mathrm{eff},s}$ & 1.623683 & 1.623682\\
  $n_{\mathrm{eff},i}$ & 1.579560 & 1.579579\\
  $(n_{g,p},n_{g,s},n_{g,i})$ & (1.7706, 1.7079, 1.8213) & (1.7705, 1.7079, 1.8206)\\
  $(D_s,D_i)$ (ps/m) & $(209.3,-168.8)$ & $(208.8,-167.3)$\\
  $\rho$ & $-0.806$ & $-0.801$\\
  $\beta_{2,i}$ (ps$^2$/m) & $-9.02$ & --\\
  Idler field share in the liquid & 0.76 & 0.76\\
  Pump intensity FWHM (THz) & 0.813 & --\\
  Purity, constant overlap (exact / first-order) & 0.788 / 0.808 & --\\
  \bottomrule
 \end{tabular}
\end{table}

\paragraph{Independent COMSOL check.}  We rebuilt the same rectangular
cores and material dispersion in a two-dimensional full-vector COMSOL
6.4 eigenmode model, using second-order electric-field elements with maximum
sizes of 50~nm in the liquid and 20~nm in Si$_3$N$_4$.  The group indices
use the same five-point stencil as femwell.  In the coordinate system defined
above, the computational rectangle is $-5.5<x<5.5~\mu$m and
$-4.0<y<5.6~\mu$m.  Silica fills the region outside the two cores;
the exterior perfect-electric-conductor condition truncates the numerical
domain and does not represent a metal in the proposed device.
Table~\ref{tab:S_liquid} compares the two models at the
nominal dimensions.  The effective indices agree to within $2\times10^{-5}$,
and the delays keep opposite signs, with $\rho=-0.801$ against $-0.806$.  The
fixed-color residual is $\Delta k_0=-95.5~\mathrm{m^{-1}}$ in COMSOL against
$14.4~\mathrm{m^{-1}}$ in femwell, a difference similar to that between the two
solvers for AlN/TiO$_2$ (Table~\ref{tab:S_femwell}).

Holding the Si$_3$N$_4$
width and gap fixed, COMSOL channel widths of 2629 and 2631~nm give residuals of
$-336.6$ and $+144.3~\mathrm{m^{-1}}$, and a secant step to 2630.4~nm gives
$+0.4~\mathrm{m^{-1}}$ with the same mesh settings.  The independent model
therefore also admits direct carrier phase matching, after a 0.4-nm width
correction; $\rho$ was evaluated only at the nominal width.

\paragraph{Outer-domain sensitivity.}  With the core geometry, vertical
limits, and mesh settings fixed, increasing the COMSOL lateral domain width
from 11 to 15 and 19~$\mu$m changes the nominal-width $\Delta k_0$ by less
than $0.6~\mathrm{m^{-1}}$ across the three domains.  The delay ratio was
evaluated at 11 and 19~$\mu$m only and changes by less than $10^{-4}$.
Separately, moving both vertical boundaries outward by 2~$\mu$m, to
$-6.0<y<7.6~\mu$m, at the 11-$\mu$m lateral width changes the
nominal-width $\Delta k_0$ by less than $0.3~\mathrm{m^{-1}}$ and $\rho$
by less than $10^{-4}$.  At the retuned channel width of 2630.4~nm, the
residual remains below $1~\mathrm{m^{-1}}$ in the nominal domain, the
19-$\mu$m-wide domain, and the vertically enlarged domain.  The femwell
lateral boundaries lie between those of the 11- and 15-$\mu$m COMSOL domains,
and its vertical limits match the nominal COMSOL domain.  The changes under
these boundary displacements are much smaller than the approximately
$110~\mathrm{m^{-1}}$ difference between the solvers, indicating that
outer-boundary placement is unlikely to be its dominant cause.  These
carrier and local-delay checks do not establish the broadband JSA or the
absolute nonlinear overlap in COMSOL; the purity and overlap values above
remain femwell results.

\FloatBarrier
\section{Reduced-model consistency and limitations}
\label{sec:S_phasecheck}

Appendix~\ref{sec:S_phaseclosure} uses a two-mode matrix whose diagonal
elements are the isolated-guide propagation constants.  To check it against
both designs, we solved the isolated main-guide pump, signal, and idler modes and
the isolated auxiliary idler mode with the solver of Appendix~\ref{sec:S_femwell},
on the mesh of the coupled structure with the other core replaced by silica.
The coupling was obtained from the calculated half-splitting and the isolated
detuning, $\kappa^2=(\Omega/2)^2-[(\beta_{iA}-\beta_{iB})/2]^2$, and $\kappa'$
from the frequency derivative of the same relation; all slopes use the
five-point stencil of Appendix~\ref{sec:S_femwell}.  Because $\kappa$ is taken from
the full-vector splitting, the product test and the values of $u$ below are
consistency checks of the two-mode reduction, not independent predictions.
For the participation, the
full-vector lower supermode was projected onto the two power-normalized
isolated modes with the mode-overlap integral, with and without correcting for
the overlap of the two isolated modes (0.04 and 0.08).
Table~\ref{tab:S_phasecheck} lists the results at the nominal colors.  With
the projected $f_A$, Eq.~\eqref{eq:S_kappa_inference} gives $\kappa$ of
2.22--2.23 and 1.98--2.01 (in units of $10^4~\mathrm{m^{-1}}$), consistent
with the splitting-based values in the table.

Both designs satisfy the sign condition of Eq.~\eqref{eq:S_phaseproduct} for
the lower branch: $\beta_*$ lies below both isolated idler modes, and the main
guide is detuned by 0.87 (AlN/TiO$_2$) and 0.71 (liquid)
times the coupling.  The participation $1/(1+u^2)$ agrees with the projection
to within 0.03.  The product test holds to within 5\% for
AlN/TiO$_2$ but only to about 22\% for the liquid design.  The
two isolated modes span the supermode almost completely: allowing for their
overlap, they carry 99.9\% and 99.7\% of its power.  Despite this near-complete
span, the sum of the two supermode propagation constants exceeds that of the
isolated modes (the trace defect in Table~\ref{tab:S_phasecheck}) by
0.03 and 0.20 times the coupling.  These observations are consistent with
errors in the reduced matrix elements, including mutual loading omitted when
Eq.~\eqref{eq:S_H} uses isolated-guide diagonal values.  The high projected
power does not exclude propagation-constant shifts from modes outside the
two-mode subspace.  The delay ratio is more sensitive, because it depends on
$u$ through $mu^2/(1+u^2)$ and on the estimated coupling dispersion
($q=-0.19$ and $-0.11$).
Equation~\eqref{eq:S_phaseconditioned} gives $-0.88$ against the
full-vector $-0.91$ for AlN/TiO$_2$, but $-0.57$ against
$-0.81$ for the liquid design, whose larger contrast, $m=4.4$,
amplifies the loading error; with $u=-0.76$ from the projected
eigenvector instead, the estimate moves to $-0.70$.  Conversely,
solving Eq.~\eqref{eq:S_targetquadratic} for the full-vector $\rho$ gives
$u=-0.89$ and $-0.80$, against $-0.87$ and $-0.71$;
for the liquid design the difference corresponds to about 8~nm of
channel width.  The relation is therefore consistent with the branch, the
detuning, and the participation of the full-vector designs, whereas the delay
ratio and all reported numbers come from the full-vector supermodes.

An independent prediction would take $\kappa$ and the diagonal elements from
overlap integrals of the isolated fields instead of from the splitting.  We
evaluated such a model only at the idler carrier: first-order non-orthogonal
coupled-mode theory with the self-coupling terms, built from the same isolated
modes.  No frequency derivatives of this model were computed, so it gives no
independent delay ratio.  It reproduces the calculated splittings to within
0.01\% (AlN/TiO$_2$) and 0.8\% (liquid),
but it places both supermodes below the full-vector values: by
466 and 462~m$^{-1}$ for the lower and upper branches of AlN/TiO$_2$,
and by 911 and 1226~m$^{-1}$ for the liquid design
(Table~\ref{tab:S_phasecheck}).  The lower-branch offsets correspond
to about 0.35~nm of AlN width and 3--4~nm of channel width (from
the fixed-color slopes of the coupled AlN/TiO$_2$ structure and of the isolated
channel), and they are comparable to or larger than the 5-mm phase-matching
acceptance ($\mathrm{sinc}^2$ falls to one half at
$|\Delta k|\simeq5.6\times10^{2}~\mathrm{m^{-1}}$).  The isolated-mode relations
can therefore guide a search, but they cannot replace the full-vector
phase-matching calculation; this is a limitation of the present analysis.

\begin{table}[h]
 \centering
 \caption{Isolated-mode consistency check of the phase-closure relation
 (femwell, nominal colors).  $\beta_*=\beta_{pA}-\beta_{sA}$ uses the isolated
 main-guide pump and signal; $\kappa$ is inferred from the calculated
 half-splitting, so the rows above the rule are consistency checks; the trace
 defect is $(\beta_{i,+}+\beta_{i,-})-(\beta_{iA}+\beta_{iB})$.  The projection
 range covers the raw and overlap-corrected weights.  The last two rows are
 from first-order non-orthogonal coupled-mode theory built only from the
 isolated fields, at the idler carrier.}
 \label{tab:S_phasecheck}
 \begin{tabular}{lll}
  \toprule
  Quantity & AlN/TiO$_2$ & FNLC/Si$_3$N$_4$\\
  \midrule
  $n_{iA}$ (isolated main guide) & 1.83881 & 1.58408\\
  $n_{iB}$ (isolated auxiliary) & 1.84039 & 1.58652\\
  $\lambda_i\beta_*/2\pi$ & 1.83288 & 1.57964\\
  $(n_{i,-},n_{i,+})$ (supermodes) & (1.83284, 1.84659) & (1.57956, 1.59230)\\
  $\Delta_A,\ \Delta_B$ ($10^4$~m$^{-1}$) & $-1.94,\ -2.46$ & $-1.46,\ -2.25$\\
  $\kappa$ ($10^4$~m$^{-1}$) & 2.24 & 2.05\\
  $\Delta_A\Delta_B/\kappa^2$ & 0.96 & 0.78\\
  Trace defect$/\kappa$ & 0.03 & 0.20\\
  $u=\Delta_A/\kappa$ & $-0.87$ & $-0.71$\\
  $f_A$: $1/(1+u^2)$ / projection & 0.57 / 0.57--0.58 & 0.66 / 0.63--0.66\\
  $r,\ m,\ q$ & $0.74,\ 3.31,\ -0.19$ & $1.02,\ 4.42,\ -0.11$\\
  $\rho$: Eq.~\eqref{eq:S_phaseconditioned} / supermodes & $-0.88$ / $-0.91$ & $-0.57$ / $-0.81$\\
  \midrule
  $\kappa$ from isolated-field overlaps ($10^4$~m$^{-1}$) & 2.23 & 2.01\\
  CMT $\beta_{i,-},\ \beta_{i,+}$ minus full vector (m$^{-1}$) & $-466, -462$ & $-911, -1226$\\
  \bottomrule
 \end{tabular}
\end{table}

\FloatBarrier

\section*{Disclosures}
The authors declare no conflicts of interest.

\section*{Data availability}
Data and analysis code underlying the results presented in this paper are not
publicly available at this time but may be obtained from the corresponding
author upon reasonable request.

\bibliographystyle{unsrt}
\bibliography{references}
\end{document}